\pdfoutput=1
\documentclass[aps,prl,preprint,superscriptaddress,longbibliography]{revtex4-1}

\RequirePackage{rotating}
\usepackage{graphicx}
\usepackage{epstopdf, epsfig}
\usepackage{amsmath}
 \usepackage{xcolor}
\usepackage{rotating}
\usepackage{adjustbox}
\usepackage{lineno}
\usepackage{subcaption}
\usepackage{comment}
\newcommand{\todo}[1]{\textcolor{red}{#1}}

\usepackage[english]{babel}

\begin{document}


\title{Retention of rising droplets in density stratification}


\author{Tracy L. Mandel}
\email[]{tracy.mandel@unh.edu}
\affiliation{Department of Mechanical Engineering, University of New Hampshire, Durham, NH 03824, USA}
\author{De Zhen Zhou}
\affiliation{Department of Applied Mathematics, University of California, Merced, CA 95343, USA}
\affiliation{Department of Physics, University of California, Merced, CA 95343, USA}
\author{Lindsay Waldrop}
\affiliation{Schmid College of Science and Technology, Chapman University, Orange, CA 92866}
\author{Maxime Theillard}
\affiliation{Department of Applied Mathematics, University of California, Merced, CA 95343, USA}
\author{Dustin Kleckner}
\affiliation{Department of Physics, University of California, Merced, CA 95343, USA}
\author{Shilpa Khatri}
\affiliation{Department of Applied Mathematics, University of California, Merced, CA 95343, USA}


\begin{abstract}
\noindent 
In this study, we present results from experiments on the retention of single oil droplets rising through a two-layer density stratification. These experiments confirm the significant slowdown observed in past literature of settling and rising particles and droplets in stratification, and are the first experiments to study single liquid droplets as opposed to solid particles. By tracking the motion of the droplets as they rise through a stratified fluid, we identify two timescales which quantitatively describe this slowdown: an entrainment timescale, and a retention timescale. The entrainment timescale is a measure of the time that a droplet spends below its upper-layer terminal velocity and relates to the length of time over which the droplet's rise is affected by entrained dense fluid. Shadowgraph experiments allow us to observe the decay of entrained fluid, and support this interpretation of the timescale. The retention time is a measure of the time that the droplet is delayed from reaching an upper threshold far from the density transition. These two timescales are interconnected by the magnitude of the slowdown relative to the upper-layer terminal velocity, $(U_u-U_{min})/U_u$. Additionally, both timescales are found to depend on the Froude and Reynolds numbers of the system, Fr $=U_u/(Nd)$ and Re $=\rho_u U_u d/\nu$. We find that both timescales are only significantly large for Fr $\lesssim1$, indicating that trapping dynamics in a relatively sharp stratification arise from a balance between drop inertia and buoyancy.  Finally, we present a theoretical formulation for the drag enhancement $\Gamma$, the ratio between the maximum stratification force and the corresponding drag force on the droplet, based on a simple force balance at the point of the velocity minimum. Using our experimental data, we find that our formulation compares well with recent theoretical and computational work by Zhang et al. [J. Fluid Mech. 875, 622-656 (2019)] on the drag enhancement on a solid sphere settling in a stratified fluid, and provides the first experimental data supporting their approach. 
\end{abstract}


\maketitle


\section{Introduction}


There are many examples of droplets, bubbles, and particles interacting with stratified fluids, including atmospheric and marine pollution \citep{turco_climate_1990}, oil spills \citep{camilli_tracking_2010, kessler_persistent_2011, socolofsky_multi-phase_2002, socolofsky_role_2005, gros_petroleum_2017}, oil seeps \cite{clark_dissolved_2000,leifer_engineered_2009}, falling leaves \citep{lam_passive_2019}, and marine snow \cite{macintyre_accumulation_1995,kindler_diffusion-limited_2010,prairie_delayed_2013}. As a result, understanding when and how stratification affects rising and settling is of significant interest within a variety of fields.

\begin{figure}
  \centering
  \centerline{\includegraphics[trim= {0 0 2.5cm 0}, clip, width=.95\textwidth]{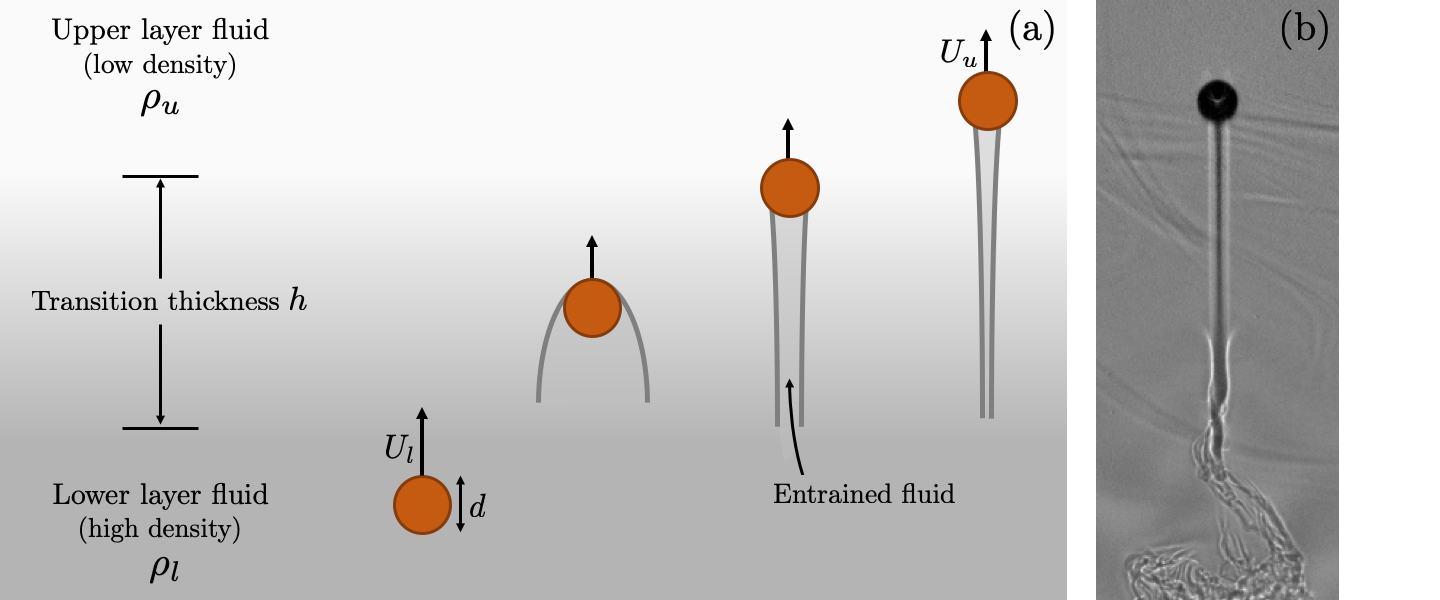}}
  \caption{(a) Schematic of a single droplet rising through the transition between two homogeneous-density layers. The droplet entrains denser fluid, decreasing its effective buoyancy as it enters the low density fluid. (b) Shadowgraph image of actual experiment showing the column of entrained fluid being dragged upward by the droplet, as well as distortion of isopycnals.}
\label{fig:general_schematic}
\end{figure}

There has been extensive prior work examining particles, drops, and bubbles rising and settling in homogeneous-density fluids, from the vortex shedding and wake dynamics of a sphere \cite{horowitz_effect_2010,auguste_path_2018} to basic statistics such as terminal rise velocity of droplets \cite{wegener_terminal_2010,baumler_drop_2011,bertakis_validated_2010}. However, density stratification adds an additional level of complexity. When a droplet rises it entrains denser fluid with it, altering the effective buoyancy of the droplet and reducing its upward speed (figure \ref{fig:general_schematic}). Furthermore, recent numerical work has suggested the presence of an additional aspect of the stratification-induced force, due to the specific structure of vorticity generated within a stratified fluid \cite{zhang_core_2019}, indicating the complex nature of this basic flow phenomenon.

Many prior experiments and numerical studies have been conducted investigating both solid and porous particles settling in stratified fluids, and will be described in detail below. However, to our knowledge, all work on the rising of droplets in stratified fluids has been computational in nature. The current study will thus focus on experimental work, to provide an expansion and validation of previous numerical studies.

The numerical simulations of Bayareh et al. \cite{bayareh_rising_2013} showed that the drag coefficient of a settling spherical drop was enhanced in linearly stratified fluids with drop Froude numbers in the range $4 \leq \text{Fr} \leq 16$ (where Fr is the ratio of the buoyancy timescale to the inertial timescale). Shaik and Ardekani found that at low Reynolds number, stratification and inertial forces both can increase the drag on a drop, and that this drag enhancement depends on the dynamic viscosity ratio between the drop and the surrounding fluid \cite{shaik_drag_2020}. Other numerical work has studied two-droplet \citep{bayareh_interaction_2016} and swarm-scale \citep{dabiri_rising_2015-1} interactions among droplets in linear stratification. These works did not involve a sharp transition between two homogeneous-density fluids, so trapping dynamics were not studied. In a two-layer stratification, the simulations of Blanchette and Shapiro \cite{blanchette_drops_2012} found that the dynamics of oil droplets in stratified fluids may have additional complexities, namely Marangoni forces. These authors found that in a sharp transition in stratification, a drop may either suddenly accelerate through the transition region or be prevented from crossing into the next layer, depending on the relative interfacial tension between the drop and the two layers. 



While experimental work on drops is limited, there is significant prior experimental work that has looked at the small-scale dynamics of rigid spheres settling in stratified fluids. The experiments of Srdic-Mitrovic et al. \cite{srdic-mitrovic_gravitational_1999} studied the gravitational settling of solid particles through a sharp two-layer stratification and found that stratification drag---that is, an increase in the drag coefficient and an associated deceleration occurring with entry into a stratified layer---was significant in only a narrow range of Reynolds numbers, $1.5 <$ Re $< 15$. Otherwise, experiments at Reynolds numbers outside of this range showed no significant change in drag as particles passed through the interface, instead behaving similarly to a particle in a homogeneous fluid. Abaid et al. \cite{abaid_internal_2004} also observed a velocity minimum for a sphere passing through a sharp transition between two homogeneous-density fluids, and in some cases reversal of the sphere's motion. Other experiments \cite{hanazaki_jets_2009,okino_velocity_2017,akiyama_unstable_2019-1} have studied the wake of spheres moving vertically in stratification, and have found that varying trailing jet structures emerge and contribute to fluid entrainment and mixing. However, these studies were conducted in linear stratification, so the effects of a sharp density transition are unknown in these regimes, and the fluid structures that emerge in a continuous stratification may be suppressed in a two-layer stratification.

 Experimental and theoretical studies of settling porous spheres have also predicted increased drag or prolonged retention times at sharp density transitions \citep{camassa_prolonged_2009,camassa_first-principle_2010,camassa_retention_2013,prairie_delayed_2013,panah_simulations_2017}, due to either diffusion of lighter fluid into the settling porous particle, or to entrainment of lighter fluid from above. However, most of these studies were limited to the Stokes regime (Re $\ll 1$). Similar studies on spheres in linear stratification \citep{yick_enhanced_2009, mehaddi_inertial_2018} have also been conducted at very low Reynolds number. Oceanic applications, such as rising bubbles or oil droplets from spills in the ocean may have Reynolds numbers ranging from intermediate to high, depending on estimates of drop diameter and drop velocity \citep{socolofsky_formation_2011-1, weber_estimating_2012}, indicating the necessity of studies to be performed beyond the low Reynolds number regime.
 
 Finally, recent theoretical and computational work by Zhang et al. \cite{zhang_core_2019} addressed the ``stratification drag" force that is commonly used as a catch-all for the contribution of stratification to changes in a sphere's motion, including increased residence time or significant slowdown \cite{srdic-mitrovic_gravitational_1999,verso_transient_2019}. The authors decomposed contributions of stratification to enhanced drag into two components: (1) entrainment forces, due to the relative buoyancy of entrained fluid dragged behind a sphere, and (2) modification of the local vorticity field due to baroclinic torque (nonalignment of density and pressure gradients), inducing an increased shear stress at the surface of the sphere, which we will refer to here as the baroclinic vorticity force. They presented a rigorous scaling of these two forces in different Prandtl, Reynolds, and Froude number regimes. While previous literature has focused heavily on the contributions of this first force due to the buoyancy effects of entrained fluid, the baroclinic vorticity contribution to stratification forces had not previously been identified, and experimental measurements corroborating this approach are lacking.
 




The present work aims to quantify and explain the retention of single oil droplets at the transition between two homogeneous-density fluids. Using laboratory experiments, we examine flow and retention for a range of drop sizes, drop densities, and ambient stratification profiles.
In section \ref{sec:ND}, we discuss the nondimensional parameters relevant to this problem. In section \ref{sec:exp_approach}, we will describe the experimental setup and measurements taken. We will discuss our results in section \ref{sec:results}, beginning with analysis of the drop's position and velocity, and then introduce two timescales relating to fluid entrainment and droplet retention. We find that these timescales are related, and that they are dependent upon the Froude and Reynolds numbers of the system, with significant fluid entrainment or drop retention only occurring for Fr $\lesssim 1$. We then develop a theoretical formulation for the drag enhancement, $\Gamma$, induced by stratification. The scaling of $\Gamma$ compares very favorably to the work of Zhang et al. \cite{zhang_core_2019} and provides the first experimental evidence supporting their approach. Finally, we will close with a discussion of the implications of this work and future directions in section \ref{sec:discussion}. 


\section{Nondimensional parameters}
\label{sec:ND}

We covered a range of the parameter space relevant to this problem, particularly in the intermediate Reynolds number regime. Table \ref{tab:exp_param} lists the parameter definitions and ranges covered in this study, which spanned 179 different droplet experiments. Non-dimensional parameters with the subscript $f$, such as Re$_f$, encompass two separate nondimensional numbers for a single drop's behavior in the upper and lower layers of ambient fluid, Re$_u$ and Re$_l$. The subscript $f$ in the given definition is thus replaced by the corresponding upper ($u$) or lower ($l$) layer quantity. The subscript $d$ represents the corresponding nondimensional number or parameter for the drop fluid properties. In these definitions, $\rho_f$ represents fluid density, $U_f$ represents the terminal drop velocity in a given homogeneous-density region, $d$ is the drop diameter, $\mu$ is the dynamic viscosity of the fluid, and $h$ is the thickness of the transition region, computed as the height encompassing 95\% of the density variation between the upper and lower layers. The buoyancy frequency, $N$, is defined as $N = \sqrt{-g/\rho_u (\partial \rho/\partial z)}$, where $\partial \rho/\partial z$ is computed as the least-squares slope of a 0.6 cm-wide region in the transition region of the density profile at rest, centered at $z=0$.

\begin{table}
  \begin{center}
\def~{\hphantom{0}}
  \begin{tabular}{cccc}
      Parameter & Symbol  & Definition  &  Range of values \\[4pt]
      Drop density & $\rho_d$ & - & 0.9375 -- 0.9927 g cm$^{-3}$ \\
      Ambient fluid density & $\rho_f$ & - & 0.9972 -- 1.117 g cm$^{-3}$ \\
      Terminal drop speed & $U_f$ & - & 0.33 -- 13.3 cm s$^{-1}$ \\
      Drop diameter & $d$ & - & 0.15 -- 0.78 cm \\
      Transition thickness & $h$ & - & 3.0 -- 9.0 cm \\
      Buoyancy frequency & $N$ & $\sqrt{-(g/\rho_u) (\partial\rho/\partial z)}$ & 3.6 -- 7.5 s$^{-1}$ \\
      Dynamic viscosity of water & $\mu_u$ & - & 0.01 g cm$^{-1}$ s$^{-1}$\\
      Dynamic viscosity of oil & $\mu_d$ & - & 0.093 -- 0.10 g cm$^{-1}$ s$^{-1}$\\
      Drop Reynolds number & Re$_{d}$ & $\rho_d U_{u} d/\mu_u$ & 5.4 -- 540 \\
      Reynolds number & Re$_f$ & $\rho_f U_f d/\mu_u$ & 5.4 -- 1060 \\
      Archimedes number & Ar$_f$ & $g(\rho_f-\rho_d)\rho_f d^3 / \mu_u^2$ & 140 -- 700,000 \\
      Froude number & Fr & $U_{u}/(N d)$ & 0.38 -- 4.2 \\
      Prandtl number & Pr & $\nu/\kappa$ & $\sim 600$ \\
      Morton number & Mo & $g\mu_u^4(\rho_f-\rho_d)/(\rho_f^2 \gamma^3)$ & $10^{-11}-10^{-12}$ \\
      Relative density & $\Delta \rho_u$ & $(\rho_u-\rho_d)/\rho_u$ & 0.0045 -- 0.061 \\
      Relative layer thickness & - & $h/d$ & 4.5 -- 52 
  \end{tabular}
  \caption{Definition and range of parameters covered in laboratory experiments. Variables with the subscript $f$ represent that quantity in either the upper layer (e.g., $\rho_u$ and $U_u$) or lower layer (e.g., $\rho_l$ and $U_l$).}
  \label{tab:exp_param}
  \end{center}
\end{table}

Following the definitions given in table \ref{tab:exp_param}, the Reynolds number (Re$_f$ or Re$_d$) represents the ratio of inertial to viscous forces. The Archimedes number (Ar$_f$) is the ratio of buoyant forces to viscous forces. The Froude number (Fr) can be thought of in several ways: (1) as the ratio of flow inertia to external gravitational forces; (2) as the ratio of the buoyancy timescale ($1/N$) to the drop motion timescale ($d/U_u$); or (3) as a ratio of the speeds at which various information about the flow is propagating, i.e. the ratio of droplet speed to an internal wave speed. In experiments, these parameters were varied by changing the drop diameter, the drop density, and the transition region thickness (which in turn changes $N$). The Prandtl number, the ratio of momentum diffusivity to salt diffusivity, remained fixed at approximately 600 for all experiments. The Morton number is used to characterize the shape of a drop or bubble in a surrounding fluid, and is a function of the interfacial tension between the droplet and surrounding fluid, $\gamma$ (on the order of 30 mN/m for our experiments). 


\section{Experimental approach}
\label{sec:exp_approach}

\subsection{Experimental setup}
\label{sec:exp_setup}

\begin{figure}
  \centerline{\includegraphics[trim={.8cm 3.5cm 8.3cm 0.5cm}, clip, width=.95\textwidth]{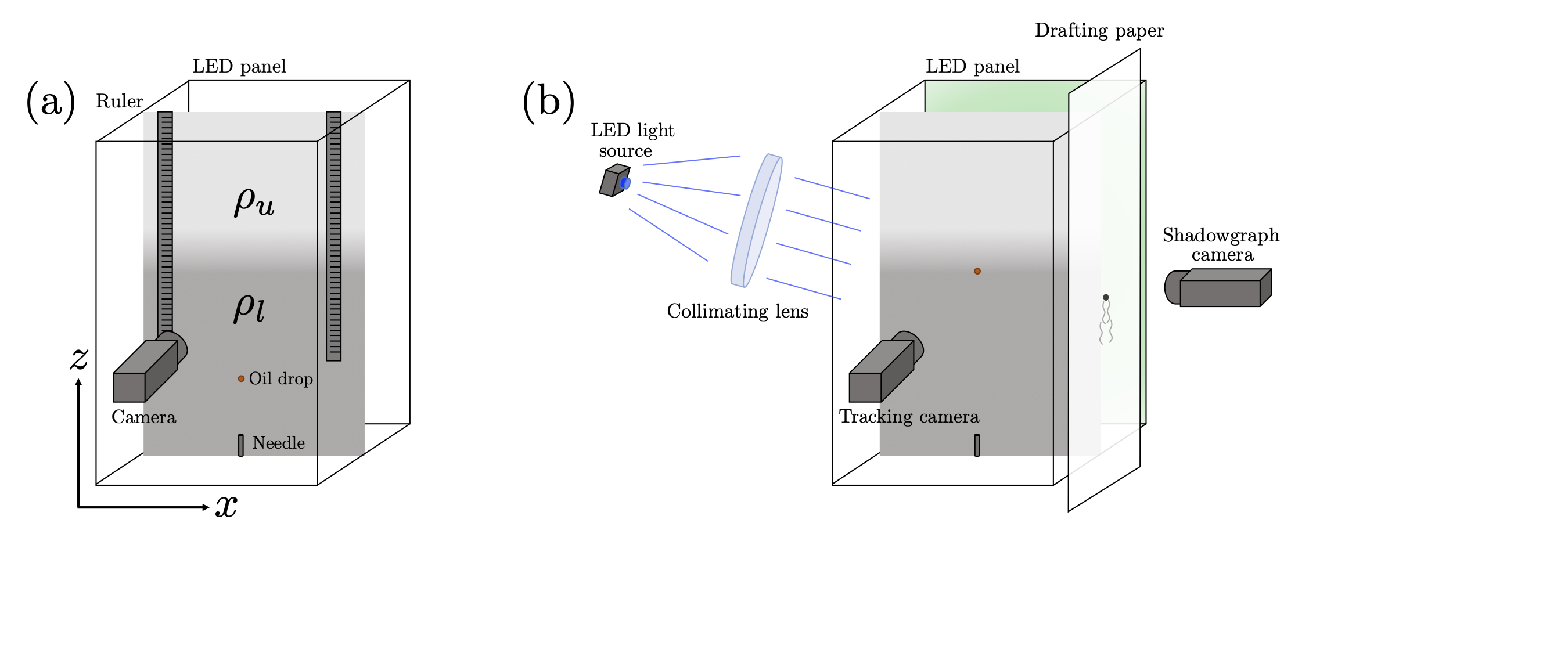}}
  \caption{Schematic of experimental setup for (a) tracking drop motion and measuring density profiles and (b) combined shadowgraph-tracking experiments.}
\label{fig:exp_schematic}
\end{figure}

Experiments were conducted in a 61 cm tall acrylic tank with a width and depth of 30.5 cm by 30.5 cm. A schematic of this tank is shown in figure \ref{fig:exp_schematic}(a). Sodium chloride (Morton Canning \& Pickling Salt) was used as the stratifying medium. Fluids for the two layers were prepared in two 35-gallon tanks with recirculating pumps, which were filled with reverse osmosis water. Salt was added to one tank and dissolved. Both tanks were left to circulate at room temperature to eliminate convection in the filled experimental tank.

\begin{figure}
  \centerline{\includegraphics[width=.8\textwidth]{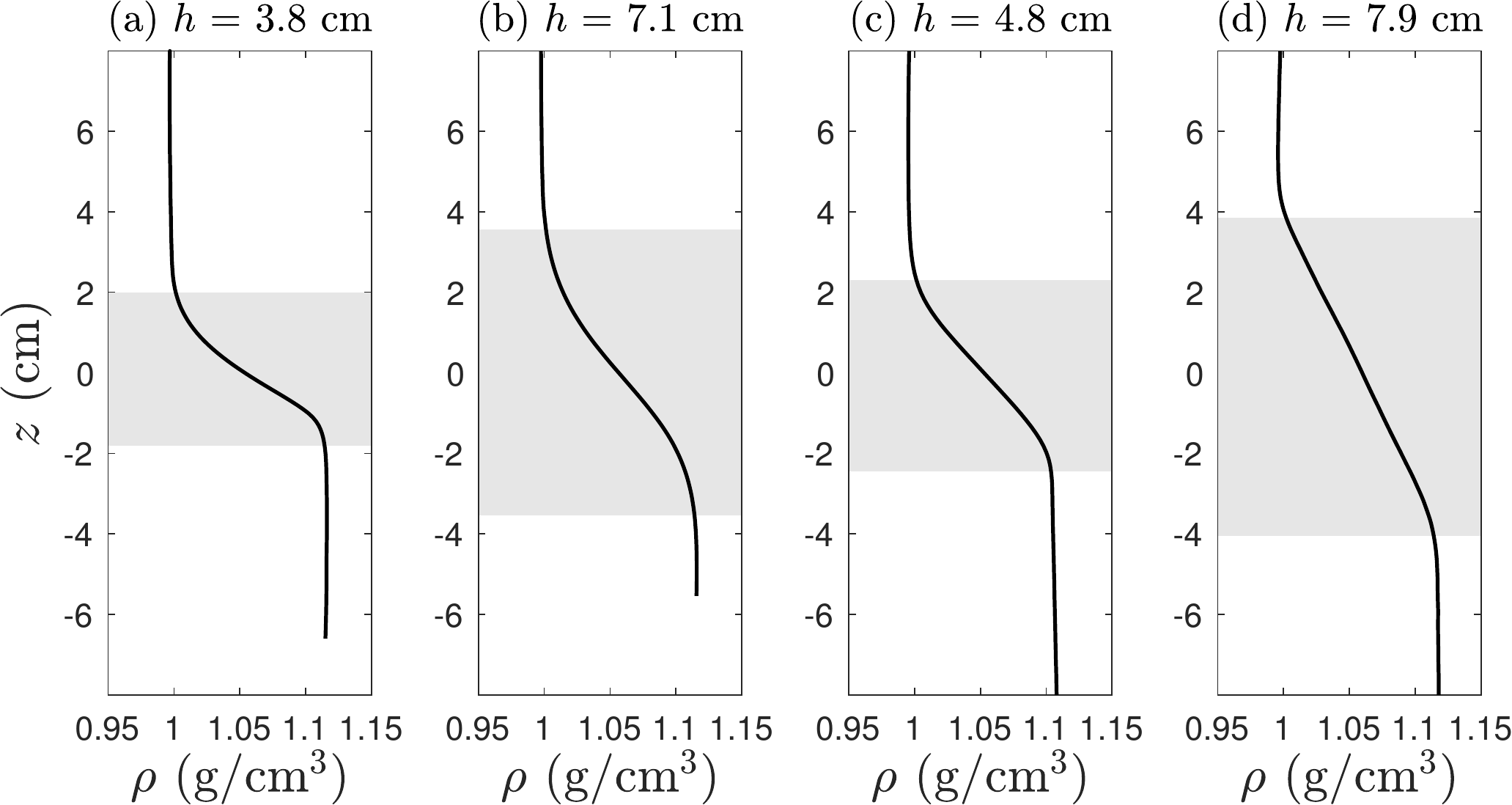}}
  \caption{Representative density profiles from experimental realizations for thinner and thicker stratified layers. Profiles were measured using synthetic Schlieren, as described in section \ref{sec:exp_meas}. The gray shading represents the transition region encompassing 95\% of the density variation between the upper and lower layers, denoted by the distance $h$. (a) The stratified layer was generated by filling the two layers directly with a sponge diffuser and then left to diffuse for 3-4 hours. (b) The layer was generated as for (a), then allowed to diffuse for 18 hours. (c,d) The layer was generated using linearly ramping pumps from fresh and salt water reservoirs in order to more precisely control transition thickness.}
\label{fig:rho_ex}
\end{figure}

The experimental tank was filled using two methods: (1) a two-layer filling method that yielded error function-type density profiles, and (2) a computer-controlled method, yielding linear density profiles in the transition region. For the first method, the tank was filled first with salt water ($\rho_l = 1.106$ to $1.117$ g/cm$^3$) and then with fresh water ($\rho_u = 0.9972$ to $0.9981$ g/cm$^3$), with a sponge float acting as a diffuser to reduce mixing between the two layers. To obtain a thin transition region (3-4 cm), the tank was allowed to sit and diffuse for 2-3 hours, until the optical distortion caused by the difference in refractive indices between the two layers had reduced. To obtain a thicker transition region (7-8 cm), the tank was allowed to diffuse another 18 hours. This two-layer filling method yielded error-function shaped density profiles, as seen in figure \ref{fig:rho_ex}(a,b). The second filling method allowed more precise control of layer thickness. For this method, two computer-controlled peristaltic pumps (New Era Pump Systems NE-9000) feeding from a fresh water bucket and a salt water bucket were linearly ramped up and down to generate a linear stratification between the upper and lower layers. These could be programmed to yield a range of transition layer thicknesses. Examples of such density profiles are shown in figure \ref{fig:rho_ex}(c,d). The gray region in these density profiles represents the layer thickness $h$, where 95\% of the density variation between the upper and lower layers occurs.

Oil droplets were composed of a mixture of 10 cSt silicone oil (Clearco Products Co.) and halocarbon oil (Sigma Life Science Halocarbon oil 27) in order to study a range of drop densities $\rho_d$, from 0.9375 to 0.9927 g/cm$^3$. Oil fluorescent tracer (Risk Reaction DFSB-K175 UV Orange) was also added for contrast. Drops were released individually by dispensing a small amount of oil using a syringe pump or handheld syringe, which fed into a 19 gauge needle inserted through a flange in the base of the tank. A waiting time of at least 15 minutes between drop releases was chosen to ensure the tank was quiescent for each experiment.

\subsection{Experimental measurements}
\label{sec:exp_meas}

Physical characteristics of the ambient fluid and oil droplets were measured prior to experiments. An Anton Paar Lovis 2000 ME microviscometer and DMA 4100 M densitometer were used to measure the viscosities and densities of the fresh water, salt water, and droplet fluid. The densitometer also provided direct measurements of the fluid's temperature.

\begin{figure}
  \centerline{\includegraphics[width=.6\textwidth]{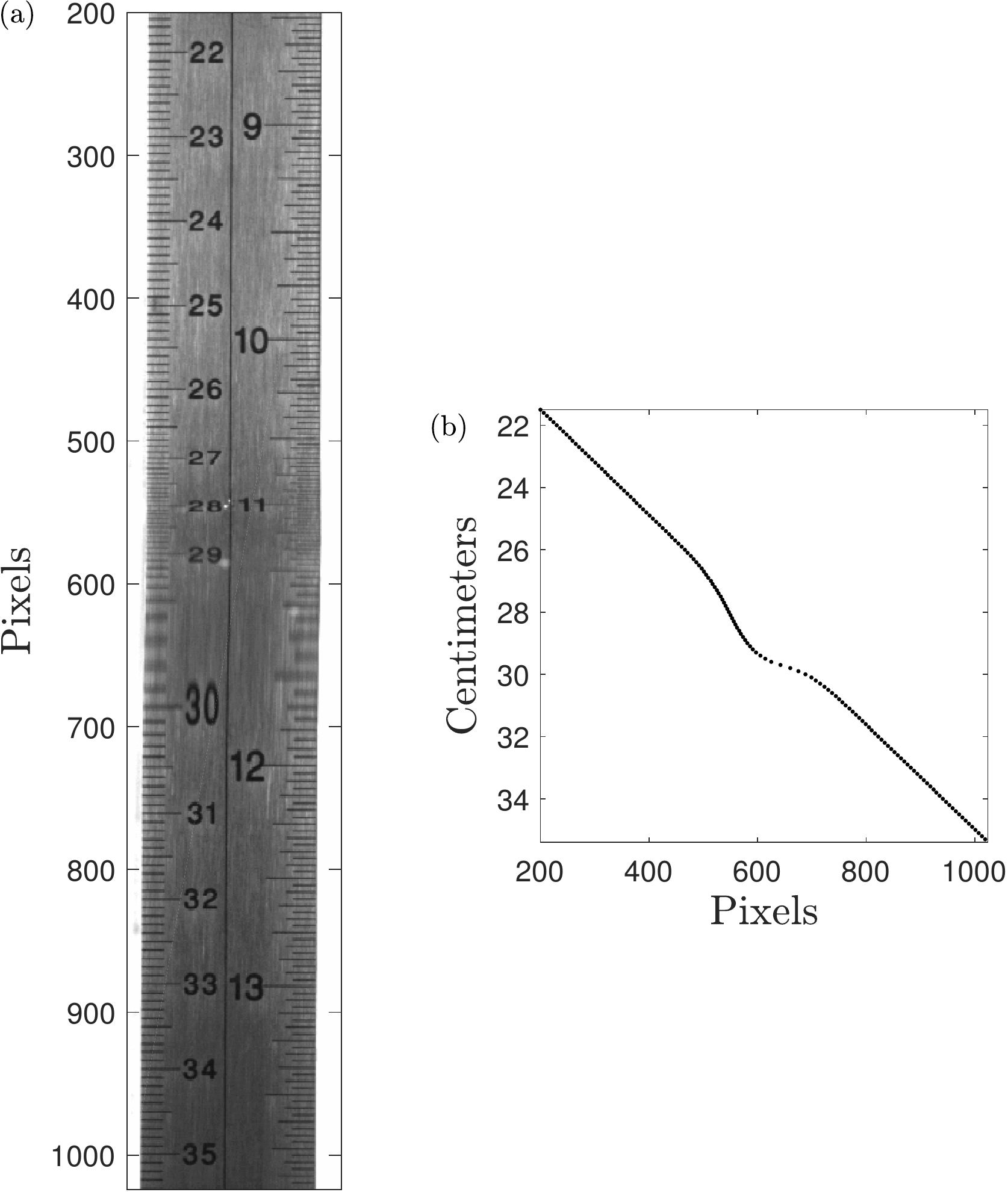}}
  \caption{(a) Example ruler image used for remapping images to remove refractive distortion. (b) Plot of centimeter reading on ruler versus pixel location.}
\label{fig:distortion_correction}
\end{figure}

During experiments, images of the injected drops were captured at 120 to 125 fps using a high-speed camera (Photron FASTCAM SA3 at 1 MP, Point Grey Grasshopper3 at 5 MP) aligned with the plane of drop motion. A panel of light emitting diodes (LEDs) was placed behind the tank, along with a diffusive screen of vellum paper between the tank and lights, to increase the contrast between drops and the background. Before each drop was released, an image was taken of the field of view, including a calibration ruler placed in line with the needle and plane of droplet motion. Because of the tank's density stratification, the refractive index encountered by a light ray changes as light passes through the tank. The images captured by the camera thus have refractive distortion. This distortion was corrected by calibrating the drop position relative to the refracted ruler image, as shown in figure \ref{fig:distortion_correction}. In some cases, the drops exhibited slight out-of-plane motion, which we estimate to contribute 1 percent or less error in measured vertical position, based on a camera distance of $\sim 1$ meter and out-of-plane motion on the order of 1 cm.

Following distortion correction and subtraction of a mean background image, drop position over time was then tracked using the Trackpy software package \citep{allan_trackpy_2016}, which uses center-of-mass detection to determine droplet position. Example tracked paths and velocities for five drops are shown in figures \ref{fig:shadowgraph1}, \ref{fig:shadowgraph2} and \ref{fig:z_vel_ex}. Instantaneous velocities were obtained following the methods of Srdic-Mitrovic et al. \cite{srdic-mitrovic_gravitational_1999}, in which a least-squares line was fit to a window of 7 points of vertical position ($\sim$0.06 sec of data) and the best-fit slope was assigned as the velocity of the center point in the window. Upper and lower layer terminal velocities, $U_u$ and $U_l$, were computed as the least-squares slope of the drop trajectory in regions with constant speed in the upper and lower layers, respectively.


Drop diameters were measured manually from an image in the lower layer, and calibrated from pixels to centimeters. For 145 cases, manually measured and calibrated diameters were verified against images of drops taken using a telecentric lens; telecentric images yielded diameters that varied on average by 0.16 mm from the other method (an average relative difference of less than 5\%), giving an estimate of the error in manual measurement and calibration.

In addition to correcting for refractive distortion, the quantified distortion of the calibration ruler was also used to determine density profiles using synthetic Schlieren \citep{sutherland_visualization_1999,dalziel_whole-field_2000}. Optical focusing and spreading of ticks on the ruler within the transition region were compared with the even spacing in the upper and lower layers, and the apparent displacement $\Delta z$ of these ticks was then converted to a density gradient following assumptions of linearity, two-dimensionality, and small incident ray angles, as described in the above references. The full density profile (such as those shown in figure \ref{fig:rho_ex}) was obtained by integrating the gradient from the known upper layer density using the equation below,

\begin{equation}
    \rho(z) = \rho_u + \frac{2 \rho_u}{g}\frac{1}{\beta L_{ruler}^2} \int_{z_u}^{z} \Delta z \hspace{.1cm} dz'
\end{equation}

\noindent where $\Delta z$ is the measured apparent displacement field (the difference between the curve shown in figure \ref{fig:distortion_correction}(b) and vertical height), and $\beta \simeq 1.88$ s$^2$ cm$^{-1}$ following \cite{sutherland_visualization_1999}. Because precise measurement of the exact distance between the ruler and the tank side wall, $L_{ruler}$, was difficult, and precision was difficult to maintain from experiment to experiment over 179 diffferent runs, this length was adjusted manually by $\pm 0.8$ cm to yield a profile whose constant upper and lower layer densities matched those measured with the hand-held densitometer.  

Finally, for 5 of the above experimental cases, shadowgraph experiments were performed to visualize the wakes of droplets. A schematic of this setup is shown in figure \ref{fig:exp_schematic}(b). Polyester drafting film (West Design Polydraw) was placed on one side of the tank, and a camera was placed facing the drafting paper so that drops and their wakes could be visualized via the focusing and defocusing of incoming light rays. A collimated light source (Thorlabs M450LP1 450 nm LED, in conjunction with an Edmund Optics 200 mm diameter, 800 mm focal length PCX condenser lens) was placed on the opposite side of the tank. The LED light source and collimating optics were set to angle downwards at about 30 degrees from the horizontal in order to avoid total internal reflection within the transition region, which would have occluded $\sim$ 1 cm of the droplet's path and also caused oversaturation in images. The images presented here thus show the projected fluid structures viewed at this angle, rather than a perfectly perpendicular view of the $x-y$ plane. A tracking camera was placed perpendicularly to the shadowgraph camera, opposite the LED backlighting panel, which was set to emit green light. Each camera was equipped with a bandpass filter (Thorlabs FELH0550 long-pass filter with a cut-on wavelength of 550 nm, and Thorlabs FES0500 short-pass filter with a cut-off wavelength of 500 nm) so that illumination from the backlighting panel and the 450 nm LED could be separated. Both tracking and shadowgraph images were synchronized using a function generator (Siglent SDG1025) that triggered the two cameras externally. Before a set of experiments, a calibration image was taken for each camera with a clear acrylic ruler in the field of view; the ruler was then moved to the edge of the tank and the tank allowed to settle for approximately 30 minutes before releasing droplets. Shadowgraph images were post-processed by correcting for optical distortion and then subtracting a background image.


\section{Results}
\label{sec:results}

We begin by discussing basic properties of the droplet motion. We will first demonstrate how droplet position and velocity vary with experimental conditions. We then discuss the terminal velocity of the droplets in the upper and lower layers, as a nondimensional predictor of this behavior will prove useful when discussing drag enhancement in stratification in section \ref{sec:dragenhancement}. In the next two subsections, we analyze the timescales over which fluid entrainment and significant droplet retention occur, discuss their dependence on the nondimensional parameters of the system, and delineate when drops are significantly retained at the transition region. We then connect these timescales to flow visualizations of the droplet's wake. Finally, we present a theoretical formulation of the drag enhancement based on our experimental data and compare this with the results of Zhang et al. \cite{zhang_core_2019}.

\subsection{Drop paths and velocities}
\label{sec:results1}

\begin{sidewaysfigure}
\centering
  \centerline{\includegraphics[trim={5cm 11cm 3.5cm 1cm}, clip, width=.9\textwidth]{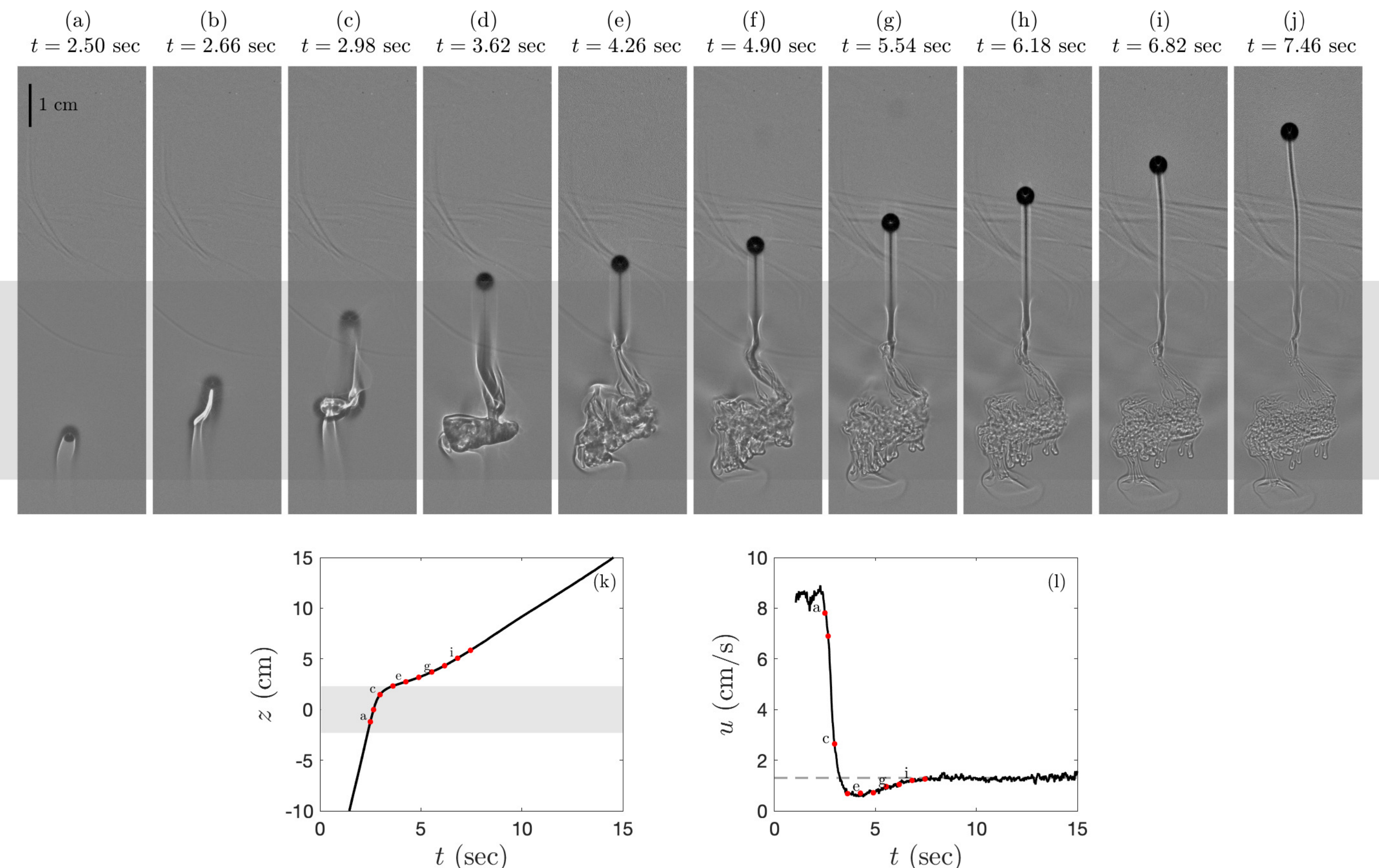}}
  \caption{(a-j) Shadowgraph images, case A. For this case, Fr $= 0.55$, $\Delta \rho_u = 0.0045$, Re$_l = 370$, Re$_u = 51.4$, and $h/d = 11.6$. Darker grey shading on the images indicates the upper and lower extents of the transition region. (k,l) Tracked position and velocities versus time. Red $(\cdot)$ symbols indicate the location of the shadowgraph snapshots. The grey region in the position versus time graph (k) indicates the extent of the transition region. See Supplemental Material at [Movie1] for a video animation of these shadowgraph images.}
\label{fig:shadowgraph1}
\end{sidewaysfigure}

\begin{sidewaysfigure}
\centering
  \centerline{\includegraphics[trim={5cm 11cm 3.5cm 1cm}, clip, width=.9\textwidth]{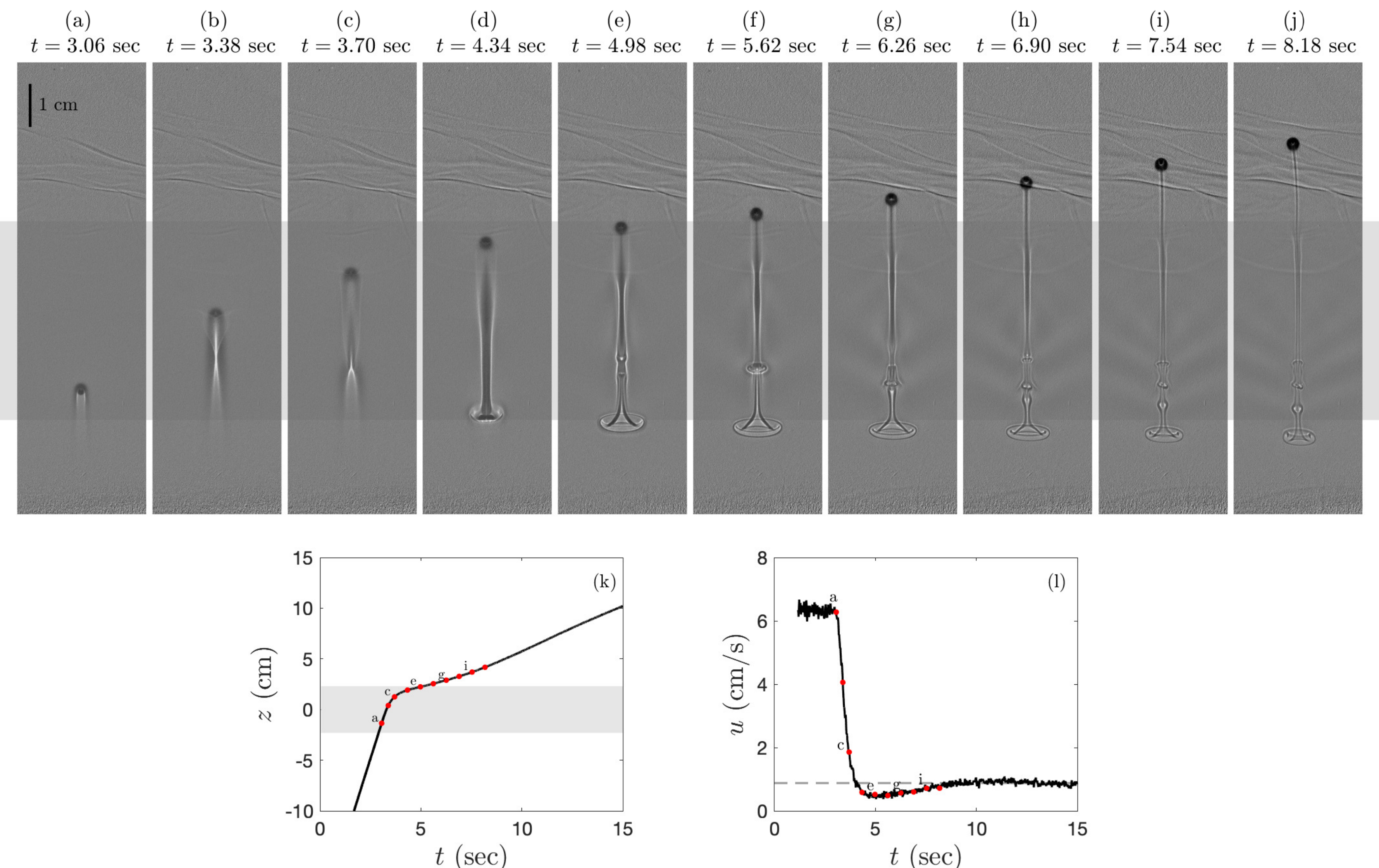}}
  \caption{(a-j) Shadowgraph images, case B. For this case, Fr $= 0.53$, $\Delta \rho_u = 0.0045$, Re$_l = 196$, Re$_u = 24.6$, and $h/d = 16.4$. See figure \ref{fig:shadowgraph1} for a full description. See Supplemental Material at [Movie2] for a video animation of these shadowgraph images.}
\label{fig:shadowgraph2}
\end{sidewaysfigure}

Shown in figures \ref{fig:shadowgraph1} and \ref{fig:shadowgraph2} are a sequence of shadowgraph images for (A) a larger (d = 0.40 cm) and (B) a smaller (d = 0.28 cm) droplet, both composed of the densest oil mixture ($\rho_d = 0.9927$ g/cm$^3$) and rising through a 4.6 cm transition region. Relative shading indicates variation in the second derivative of density \citep{settles_schlieren_2012}. As the drops exit the transition region, an internal wave field is generated. Also shown in each figure are the corresponding vertical drop position, $z$, and velocity, $u$, as a function of time, $t$, for each set of shadowgraph images. The droplets slow as they pass through the transition region (snapshots a-c in both sets of shadowgraph images), and reach a velocity minimum (snapshots e-f) just above the transition region. The drop then eventually regains speed (snapshots g-i), asymptoting to its upper-layer terminal velocity $U_u$, indicated as the dashed grey line in the velocity plots, by about snapshot j. A complex wake structure can also be observed in the shadowgraphs, which will be discussed in detail in section \ref{sec:shadowgraph_entrainment}.

\begin{figure}
  \centerline{\includegraphics[trim={0 0 0 0cm}, clip, width=\textwidth]{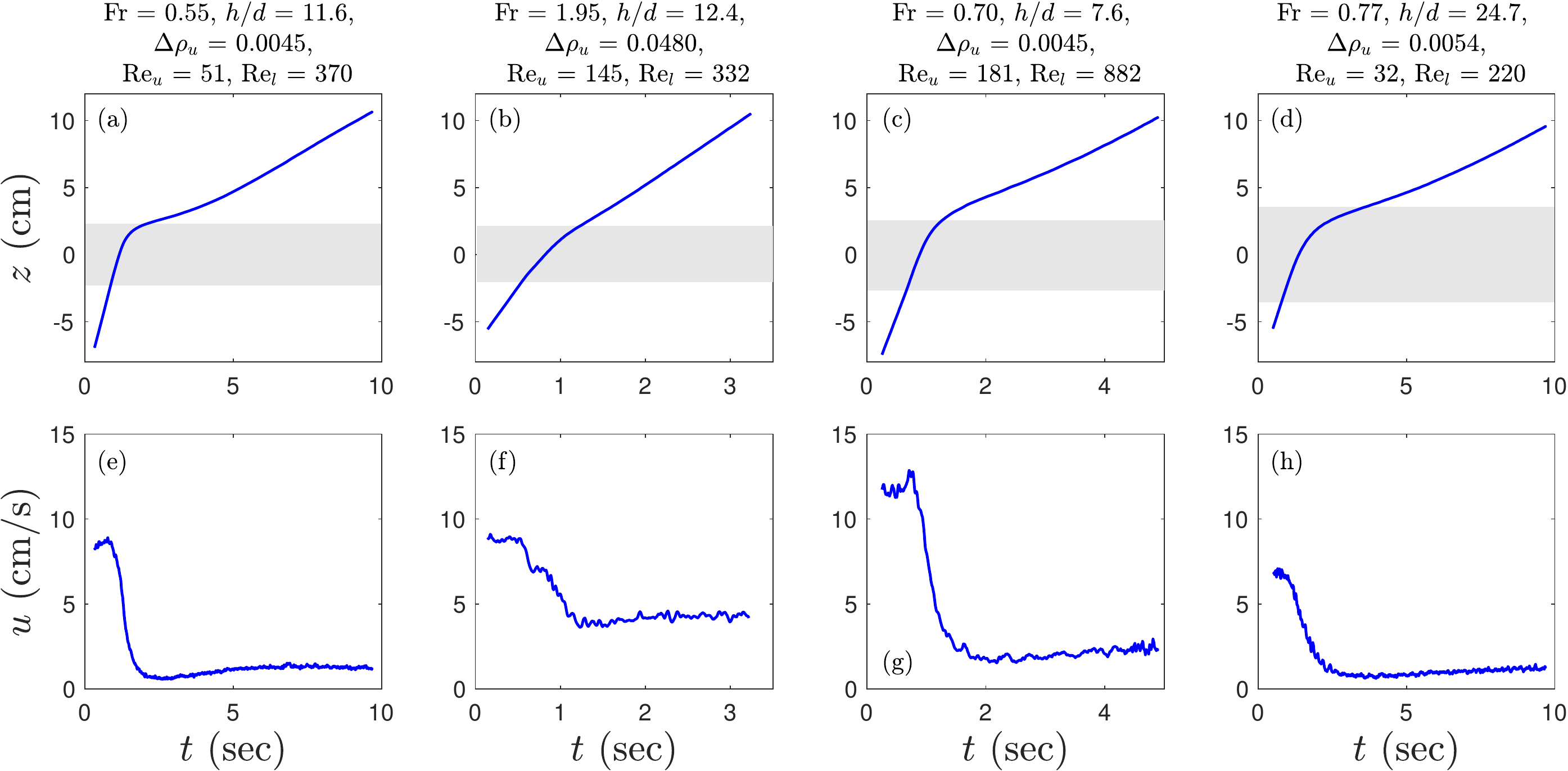}}
  \caption{Vertical positions (a-d) and instantaneous velocities (e-h) versus time for four experiments. The shaded area in (a-d) represents the transition region. (a,e) Tracking data from Shadowgraph A: a relatively small, dense reference droplet passing through a 4.6 cm transition region; 
  (b,f) A less dense droplet; 
  (c,g) A larger droplet; 
  (d,h) A small, dense droplet passing through a 7.1 cm transition region.}
\label{fig:z_vel_ex}
\end{figure}

To demonstrate how varying different experimental parameters affect drop motion, figure \ref{fig:z_vel_ex} shows sample drop paths and velocities over time for four example experimental droplet cases. The first case, shown in figure \ref{fig:z_vel_ex}(a,e), is the droplet shown in figure \ref{fig:shadowgraph2}. We use this small, dense droplet passing through a 4.6 cm transition region as a reference case for comparison with: (b,f) a similarly-sized, lighter droplet in similar ambient stratification; (c,g) a larger, dense droplet in similar stratification; and (d,h) a small, dense droplet passing through a thicker transition region of 7.1 cm. It can be seen that lighter droplets have significantly higher terminal velocities in the upper layer, and that small, dense droplets take significantly longer to traverse the field of view of the camera. The denser drops (e,g,h) also remain at a speed lower than their upper layer terminal velocity $U_u$ for an extended period of time, indicating that entrained ambient fluid plays a role in delaying the drop's upward motion. Analysis of these delays will be presented in section \ref{sec:timescales}.

\subsection{Terminal velocity}
\label{sec:terminalvelocity}

\begin{figure}
  \centerline{\includegraphics[width=.99\textwidth]{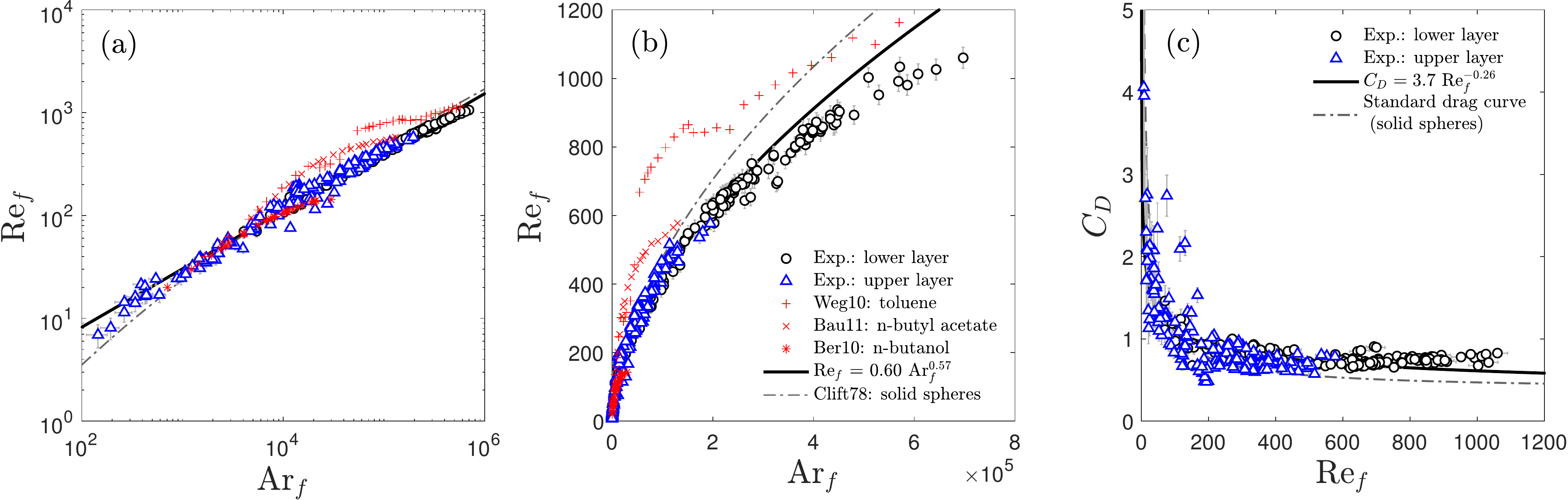}}
  \caption{Reynolds number versus Archimedes number for terminal drop behavior in the upper and lower layers, shown on (a) logarithmic and (b) linear scale. (a,b) Experimental data from this study are shown as black circles and blue triangles. A power law relationship was fit to these experimental data, with R$^2$ = 0.99, shown as the solid black line. Also plotted are terminal velocity data from studies with other droplets of varying interfacial tension in water \citep{wegener_terminal_2010,baumler_drop_2011,bertakis_validated_2010}, represented by red markers. The empirical equation for the terminal velocity of solid spheres from \cite{clift_bubbles_2005} is shown as the grey dot-dashed line. MAPE for the best fit is 16\%; MAPE between data and the relationship for solid spheres is 29\%. (c) Drag coefficient in the upper and lower homogeneous layers. A best-fit to the data is $C_D = 3.7$Re$_f^{-0.26}$, with $R^2 = 0.65$. MAPE for this best fit is 29\%; MAPE between data and the standard drag curve is 48\%. Observed drag coefficients at higher Reynolds number are generally higher than the standard drag curve for solid spheres.}
\label{fig:Ar_vs_Re}
\end{figure}

Droplets reach a constant, terminal velocity in both the upper and lower layers. This terminal velocity is governed by buoyancy, viscosity, and inertia. The drop's Archimedes number in the upper and lower layer is plotted versus its Reynolds number in each layer in figures \ref{fig:Ar_vs_Re}(a,b). As noted in table \ref{tab:exp_param}, the definitions of these two nondimensional numbers is as follows:

\begin{equation}
    \text{Ar}_f = g \frac{(\rho_f - \rho_d) \rho_f d^3}{\mu_u^2}
\end{equation}

\begin{equation}
    \text{Re}_f = \frac{\rho_f U_f d}{\mu_u}
\end{equation}

\noindent where the Archimedes number is the ratio of buoyant to viscous forces and the Reynolds number is the ratio of inertial to viscous forces. As before, the subscript $f$ represents either the upper ($u$) or lower ($l$) layer property, and $U_f$ represents the terminal drop speed in that corresponding layer. Each drop thus yields two data points of Ar$_f$ and Re$_f$. Measurements of Ar$_l$ and Re$_l$ are represented as black circles, while values of Ar$_u$ and Re$_u$ are shown as blue triangles.

A power-law fit was found to describe the relationship between these two parameters from experimental measurements:

\begin{equation}
    \text{Re}_f = 0.60 \text{Ar}_f ^ {0.57}
\label{eq:Re_Ar}
\end{equation}

\noindent with $R^2 = 0.99$. This compares reasonably well with the relationship for solid spheres in the range $435 <$ Ar $\leq 1.16 \times 10^7$ and $12.2 < $ Re $ \leq 6.35 \times 10^3$ \cite{clift_bubbles_2005}:

\begin{equation}
    \log_{10}{\text{Re}} = -1.81391 + 1.34671 W - 0.12427 W^2 + 0.006344 W^3,
\end{equation}

\noindent where $W = \log_{10}(4/3$Ar), shown as the dot-dashed line in figures \ref{fig:Ar_vs_Re}(a,b). It should be noted that this empirical equation for terminal velocity is for a solid sphere, not a deformable liquid drop. 


In figure \ref{fig:Ar_vs_Re}, as well as later figures, we assessed deviation between data and the empirical fits using the mean absolute percent error (MAPE), defined as

\begin{equation}
    \text{MAPE} = \frac{100\%}{M}\sum_{i=1}^M \left| \frac{y_{i,data}-y_{i,fit}}{y_{i,data}} \right|
    \label{eq:mape}
\end{equation}

\noindent where $y$ is the variable on the vertical axis and $M$ is the number of experimental cases ($M = 179$).

Also shown in figures \ref{fig:Ar_vs_Re}(a,b) are terminal velocities from experiments and numerical simulations of drops rising in homogeneous-density water from several other studies with varying Morton number. The Morton number (table \ref{tab:exp_param}) describes the importance of gravity and viscosity relative to interfacial tension and inertia, and is used to characterize the shape of drops and bubbles. Shown are velocities for toluene droplets (Mo = $1.95\times 10^{-11}$) \cite{wegener_terminal_2010}, \textit{n}-butanol acetate droplets (Mo = $2.87 \times 10^{-10}$) \cite{baumler_drop_2011}, and \textit{n}-butanol droplets (Mo = $1.23\times 10^{-6}$) \cite{bertakis_validated_2010}. For some of these cases in the literature, the drop aspect ratio (the ratio of vertical drop diameter to horizontal drop diameter) was much less than 1 due to a higher Morton number, which explains some of the divergence from our experimental data, which we estimate to have a Morton number on the order of $10^{-11}$ to $10^{-12}$. Although there exist many more studies that compute a terminal velocity relation for droplets or nonspherical solids in homogeneous fluid \citep[e.g.][]{wallis_terminal_1974,ganguly_prediction_1990}, the majority are for low Reynolds number flows and not applicable here.

Figure \ref{fig:Ar_vs_Re}(c) also shows the approximate drag coefficient of drops in each homogeneous-density layer, computed as

\begin{equation}
    C_D = \frac{4}{3} \frac{g (\rho_f - \rho_d)d}{\rho_f U_f^2}
\end{equation}

\noindent following \cite{clift_bubbles_2005}. The observed drag coefficient tends to be higher than that predicted by the standard drag curve for solid spheres \citep[Table 5.2]{clift_bubbles_2005}.

In summary, for drops in the parameter space studied here, if the viscosity and density of the ambient fluid, as well as density and diameter of the drop are known, then the drop's terminal speed can be predicted with reasonable accuracy using the relation given in equation \ref{eq:Re_Ar}. This terminal velocity will be used later in section \ref{sec:dragenhancement} for our theoretical formulation of drag enhancement in stratification. 

\subsection{Entrainment and retention timescales}
\label{sec:timescales}

As shown in figures \ref{fig:shadowgraph1}, \ref{fig:shadowgraph2}, and \ref{fig:z_vel_ex}, some droplets experienced a significant slowdown as they passed through the transition region. Previous work has shown that this slowdown is in part due to the entrainment of denser fluid \cite{srdic-mitrovic_gravitational_1999}. It is thus useful to look at metrics of drop retention that measure both the length of time that denser fluid is entrained, as well as the length of time that the drop is physically retained in the transition layer.

\begin{figure}
  \centerline{\includegraphics[width=.8\textwidth]{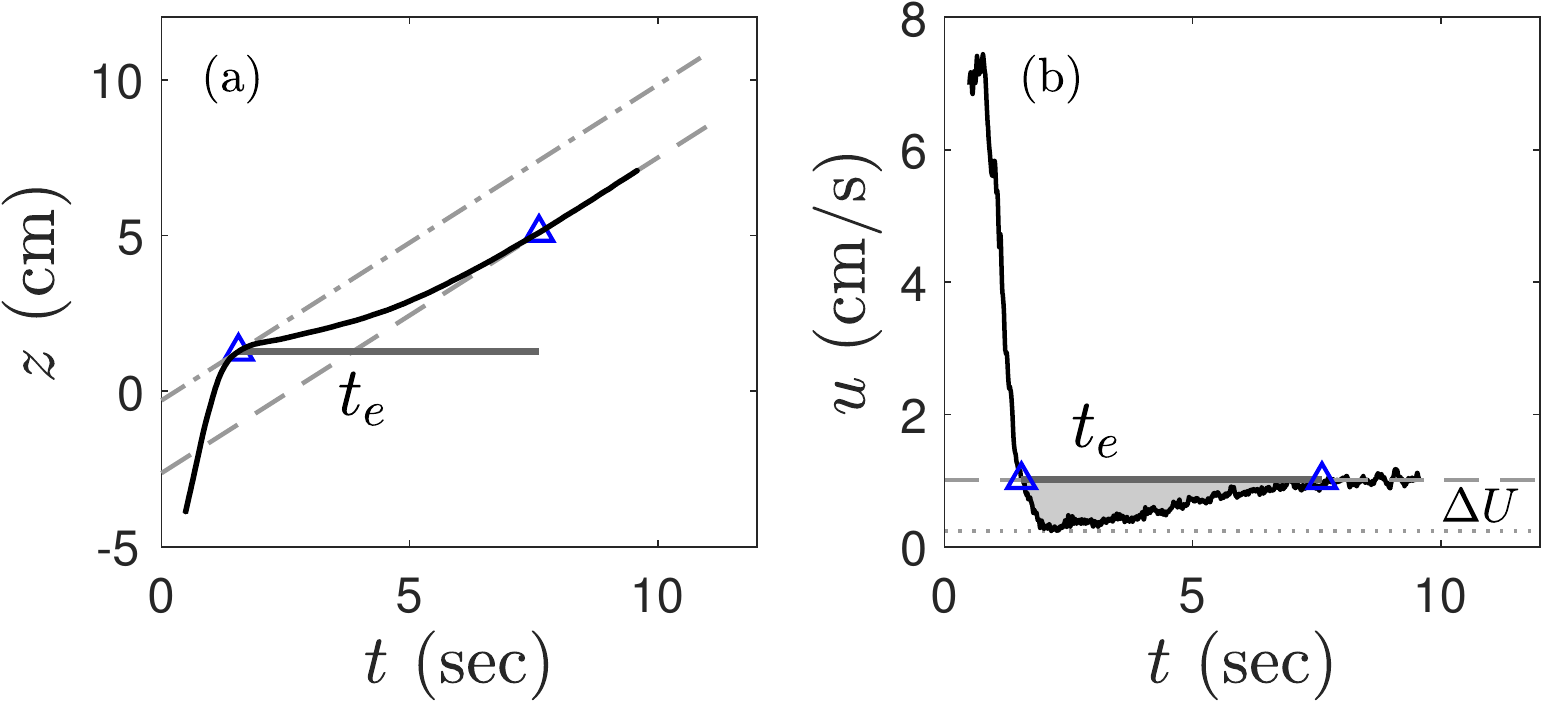}}
  \caption{Definition of drop entrainment time, $t_e$, using (a) vertical drop position and (b) drop velocity. The black line represents the experimental measurement. The grey dashed line shows $U_u$, either as (a) the slope of the position or (b) as a velocity threshold. The grey dot-dashed line in (a) is a tangent line with a slope of $U_u$, used to find the point where the drop's velocity first reaches this value. In our definition, the entrainment time is the length of time that the drop's velocity is less than the upper layer terminal velocity. It is computed as the time between when the drop first slows to $U_u$ (first blue triangle in both figures) and when the drop again speeds up to this velocity (second blue triangle in both figures). The point where the drop has asymptotically reached its upper-layer velocity is chosen as the point at which the Euclidean distance between the drop position, represented by the black line in (a), and the grey dashed line in (a) is less than 5\% of the drop diameter. The difference $\Delta U$ between the minimum velocity $U_{\text{min}}$ and $U_u$ is shown in (b).}
\label{fig:t_e}
\end{figure}

We will consider the first metric to be the entrainment time, $t_e$. This timescale is demonstrated in figure \ref{fig:t_e}, using both vertical drop position and drop velocity. The entrainment time measures the amount of time the drop spends below its upper layer terminal velocity $U_u$, i.e., the length of time over which entrained denser fluid is appreciably slowing the droplet's motion. It is computed as the time between when the drop first slows to $U_u$ (followed by significant further slowdown), and when the drop again speeds up to this upper-layer terminal velocity. The point where the drop has asymptotically reached $U_u$ again is chosen as the point at which the distance between the drop position (the black line in figure \ref{fig:t_e}(a)) and a line representing the upper layer terminal speed (the grey dashed line in figure \ref{fig:t_e}(a)) is less than 5 percent of the drop diameter. This timescale is very similar to the delayed settling time (DST) used by Prairie et al. \cite{prairie_delayed_2013}.

\begin{figure}
  \centerline{\includegraphics[width=.8\textwidth]{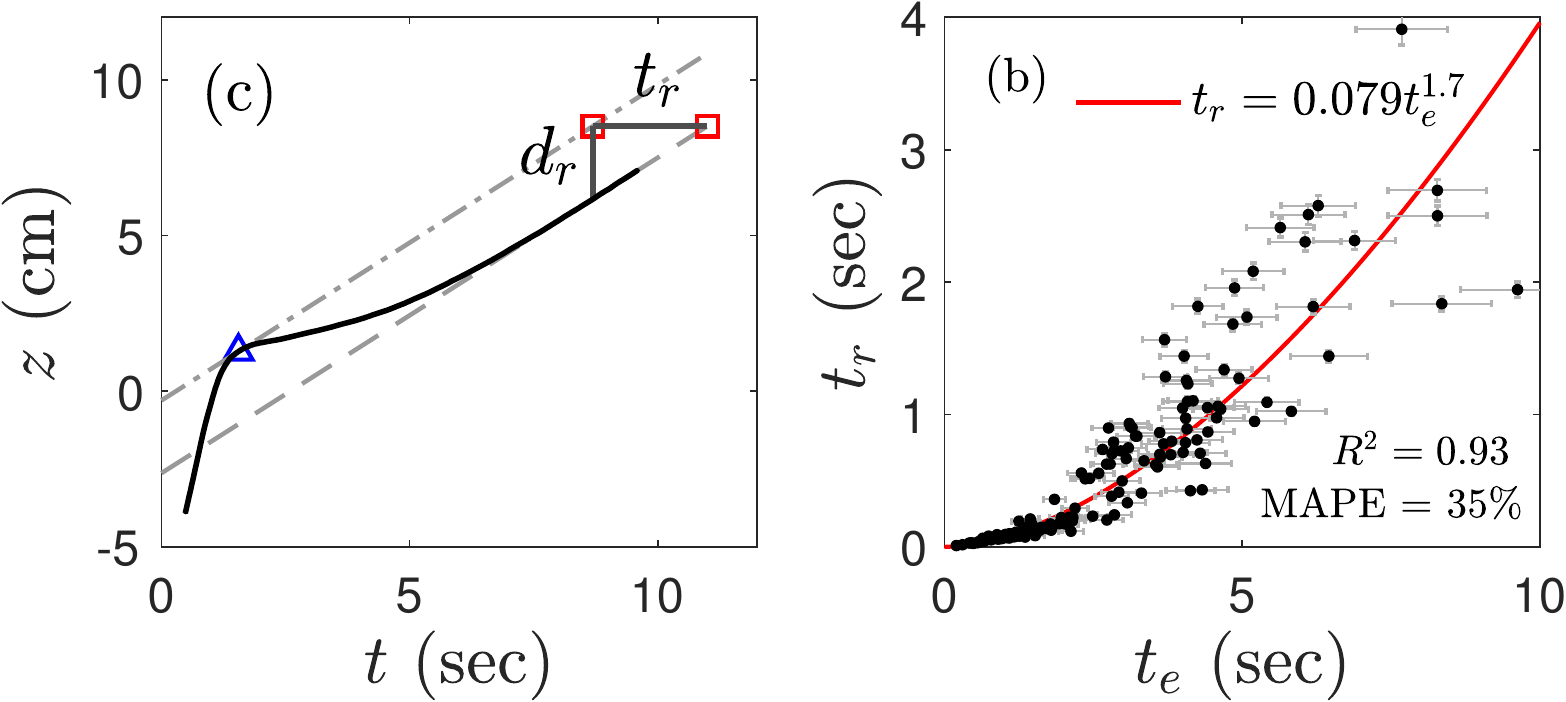}}
  \caption{(a) Definition of drop retention time $t_{r}$. $t_{r}$ is computed as the time between when the drop actually passes an upper threshold, and when it would pass the upper threshold if it hadn't slowed down further after it first slowed to $U_u$ (i.e., if the drop had instead followed the dot-dashed path after the blue triangle). The retention distance $d_r$ is also shown. (b) Comparison of $t_{r}$ with $t_e$. The best-fit power law is $t_r = 0.079 t_e^{1.7}$ with $R^2 = 0.93$. The mean absolute percent error (MAPE) is 35\%, as defined in equation \ref{eq:mape}.}
\label{fig:t_r}
\end{figure}

The second metric is a retention time, $t_r$, the time that the droplet is retained in the transition layer. This time is shown in figure \ref{fig:t_r}(a), and is computed as the time between when the drop actually passes an upper threshold ($z \approx 10$ cm, where $z=0$ is the center of the transition region), and when it would have passed the upper threshold if it had not slowed down once it first reached $U_u$. Figure \ref{fig:t_r}(b) compares values of $t_e$ and $t_r$, which follow a power-law relationship of $t_r = 0.079 t_e^{1.7}$ with an $R^2$ value of 0.93. There is obviously a strong relationship between these two metrics, with droplets that have a long entrainment timescale also being retained the longest.

\begin{figure}
  \centerline{\includegraphics[width=0.8\textwidth]{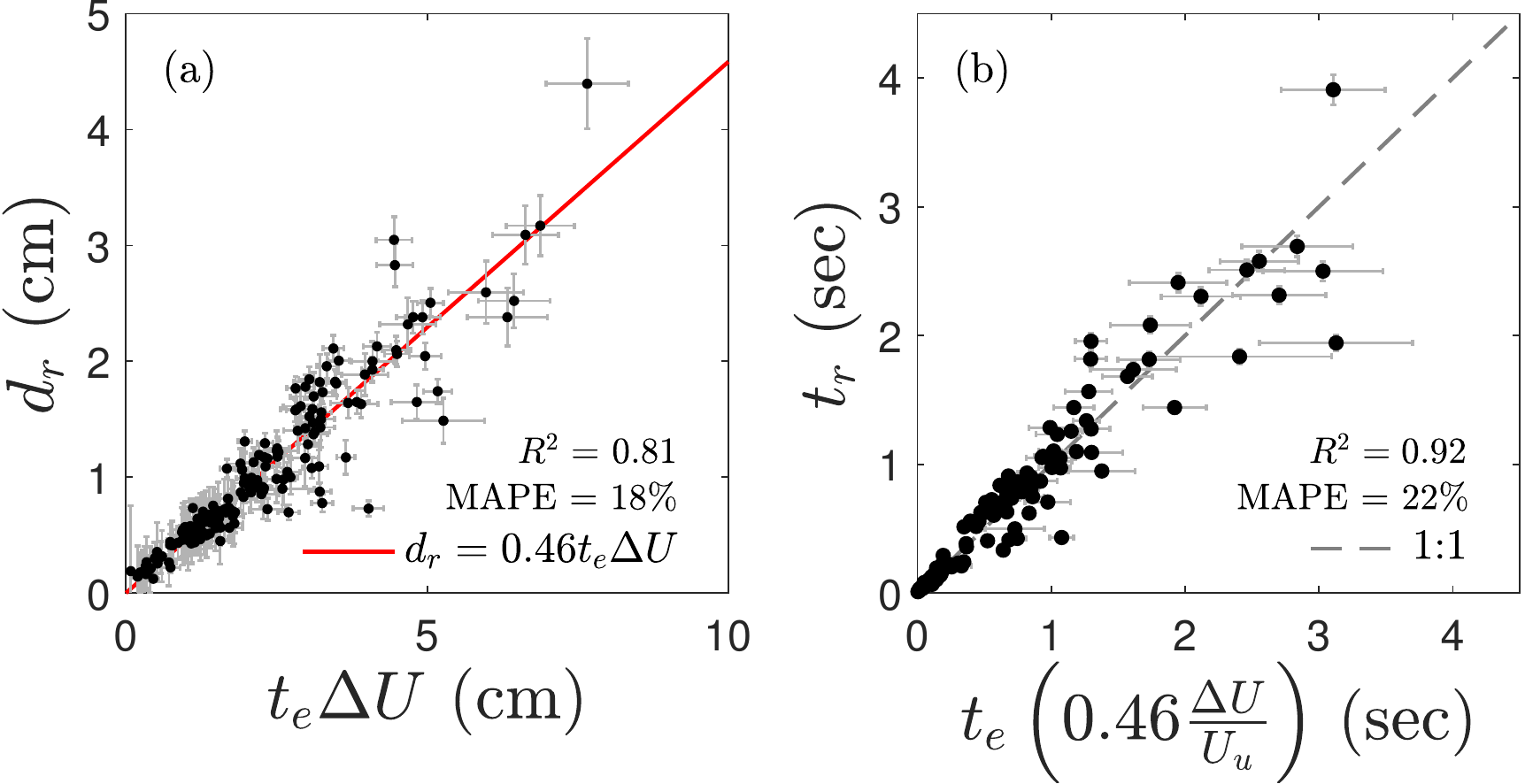}}
  \caption{(a) Best-fit for the relationship in equation \ref{eq:dr_te}, $d_r = 0.46 t_e \Delta U$, with $R^2 = 0.81$. (b) Equation \ref{eq:tr_te_eq}, with a coefficient of determination, $R^2 = 0.92$. The dashed grey line indicates a 1:1 relationship. The mean absolute percent error (MAPE) is given for both (a) and (b), as defined in equation \ref{eq:mape}.} 
\label{fig:tr_te_relationship}
\end{figure}

The retention time can also be expressed in terms of the retention distance, $d_r$, as shown in figure \ref{fig:t_r}(a), and the upper layer terminal velocity:

\begin{equation}
    t_r = \frac{d_r}{U_u}
\label{eq:tr_ud}
\end{equation}

\noindent and $d_r$ can be scaled as

\begin{equation}
    d_r \sim t_e \Delta U,
\label{eq:dr_te}
\end{equation}

\noindent where $\Delta U = U_u - U_{\text{min}}$ (figure \ref{fig:t_e}(b)).  We found a best-fit relation of $d_r = 0.46 t_e \Delta U$, with an $R^2$ value of 0.81, as shown in figure \ref{fig:tr_te_relationship}(a). This expression can then be substituted into equation \ref{eq:tr_ud}, yielding









\begin{equation}
    t_r = t_e \left(0.46 \frac{\Delta U}{U_u} \right).
\label{eq:tr_te_eq}
\end{equation}

\noindent The retention time is, therefore, the entrainment time multiplied by a constant, $c = 0.46$, and by a factor representing the relative magnitude of the drop's slowdown, $\Delta U/U_u$. This relationship is shown in figure \ref{fig:tr_te_relationship}(b), with an $R^2$ value of 0.92. The physical retention of a drop at the transition region is thus a function of the length of time that denser fluid is appreciably entrained, $t_e$, and the relative drop slowdown, $\Delta U/U_u$, which may be due to both the amount of dense fluid entrained as well as any contributions of baroclinic vorticity generation to the stratification force.

\begin{figure}
  \centerline{\includegraphics[trim={2.2cm 0 .9cm 0}, clip, width=\textwidth]{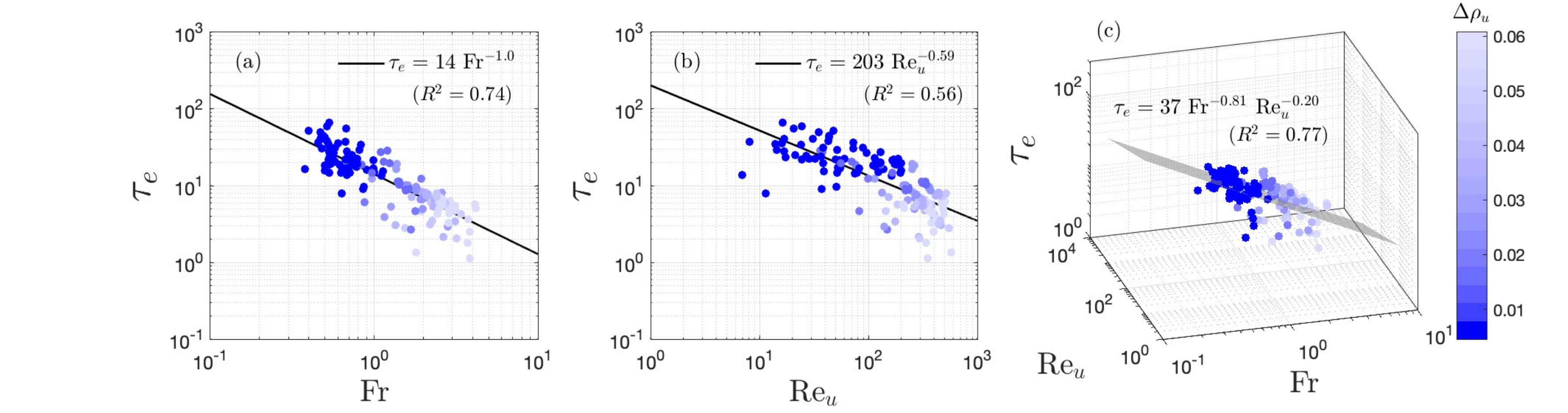}}
  \caption{Nondimensional drop entrainment time, $\tau_e$, versus Fr and Re$_u$. Shading indicates the drop's relative density to the upper layer, $\Delta \rho_u$.} 
\label{fig:taue_3D}
\end{figure}

\begin{figure}
  \centerline{\includegraphics[trim={2.2cm 0 .9cm 0}, clip, width=\textwidth]{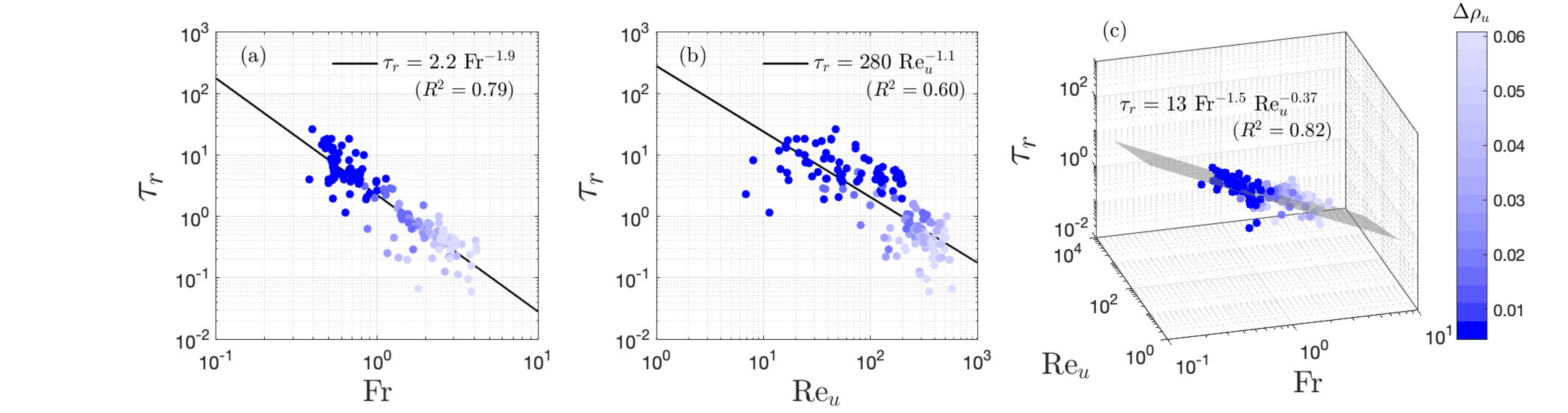}}
  \caption{Nondimensional drop retention time, $\tau_r$, versus Fr and Re$_u$. Shading indicates the drop's relative density to the upper layer, $\Delta \rho_u$.} 
\label{fig:taur_3D}
\end{figure}

In order to understand how environmental conditions affect these timescales, the retention and entrainment times were then compared against nondimensional parameters governing the drop's rise. Figures \ref{fig:taue_3D} and \ref{fig:taur_3D} show nondimensional entrainment time, $\tau_e = t_e N$, and nondimensional retention time, $\tau_r = t_r N$, as a function of the Froude number and Reynolds number in the upper layer. These timescales are more strongly correlated with the Froude number than the Reynolds number, indicating the importance of stratification in drop retention.  Other nondimensional numbers, including $Re_l$, $h/d$, and a lower-layer Froude number, were also compared but did not yield significant collapse of the data and so are not presented here.

\begin{figure}
  \centerline{\includegraphics[width=.9\textwidth]{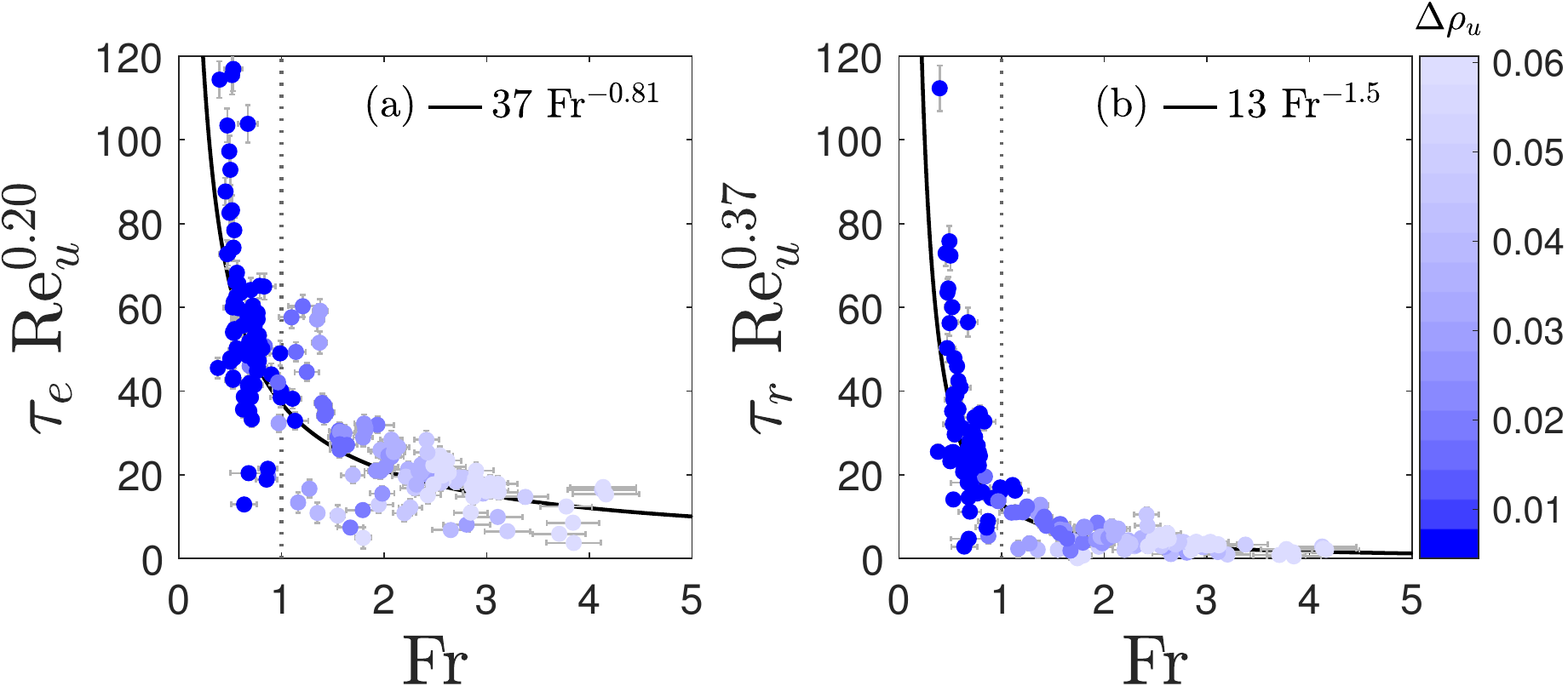}}
  \caption{Nondimensional entrainment and retention times versus Reynolds and Froude number. Shading indicates the drop's relative density to the upper layer. Best-fit power laws are shown as the black line in (a) and (b), which are simply rearrangements of the equations in figures \ref{fig:taue_3D}(c) and \ref{fig:taur_3D}(c).}
\label{fig:tau_vs_Fr}
\end{figure}

Nondimensional entrainment and retention times, with their associated Reynolds scaling from figures \ref{fig:taue_3D}(c) and \ref{fig:taur_3D}(c), are plotted against Fr in figure \ref{fig:tau_vs_Fr}. Both of these temporal metrics have a power law relationship with Froude number, with larger nondimensional retention and entrainment times occurring for small Fr. A delineation can be drawn at Fr $\approx 1$, and we can split droplet behavior into two regimes:
\begin{enumerate}
    \item For Fr $\geq 1$, the drop is rising on a timescale less than the buoyancy timescale, and the drop behaves as if it were in a homogeneous fluid. Retention times are close to zero, as inertial forces dominate over buoyant forces.
    \item For Fr $<1$, significant drop retention is observed. As the buoyancy timescale ($1/N$) becomes less than the inertial timescale ($d/U_u$), the drop is more significantly retained in the transition region. This corresponds to buoyancy forces being strong enough to counter the drop's inertial forces. Conceptually, one may also consider that for Fr $< 1$, the drop's velocity is less than a characteristic internal wave velocity, $N d$, and the drop does not have the kinetic energy required to ``punch through" the transition region.
\end{enumerate}

These results demonstrate that in the regime covered by our experiments, the dynamics of droplet trapping are primarily governed by a balance between buoyancy and inertia, and that the Froude number is the most important parameter to consider when predicting trapping or slowdown in a two-layer stratification.


We observed a significant slowdown for a wide range of Reynolds number, with $\tau_e > 10$ observed for Re$_u$ up to 300. Significant delay of this magnitude was only observed for Fr $\lesssim 1$ and and $\Delta \rho_u \lesssim 0.035$. Within existing literature that has examined deceleration and retention of particles and solid spheres in linear or sharp stratification, the parameter regime in which a significant delay occurs has not been well quantified. Srdic-Mitrovic et al. \cite{srdic-mitrovic_gravitational_1999} only observed significant slowdown for Reynolds numbers (based on the velocity on \textit{entry} into the stratified layer) between 1.5 and 15. The numerical simulations of Torres et al. \cite{torres_flow_2000} at intermediate Reynolds number found drag to strongly increase with Fr$^{-1}$ for Fr $<20$; however, it was found that this increase in drag was due to a rear buoyant jet that persists in a continuous stratification, and may not be applicable to the relatively sharp two-layer stratification studied here. Other work, including Yick et al. \citep{yick_enhanced_2009}, found that enhanced drag scales with Ri$^{0.51}$ in the very small Reynolds number regime, where the Richardson number Ri $=\frac{N}{\nu/d^2} \frac{1}{\text{Fr}}$. Recent work by Zhang et al. \cite{zhang_core_2019} contributed the most exhaustive study of the parameter space to date, and delineated the dominant mechanisms of drag enhancement within different regimes. Their modeling and numerical results have not been experimentally verified, and did not quantify delay in terms of time scales, an aspect of this increased stratification drag that is particularly relevant in environmental applications. We will discuss the applicability of our results to their work on drag enhancement in more detail in section \ref{sec:dragenhancement}. 

\subsection{Fluid entrainment}
\label{sec:shadowgraph_entrainment}

Above, we have given a conceptual model of how stratification ultimately affects the residence time of droplets at a density transition. Here, we discuss visualizations of the wakes of the droplets first shown in figures \ref{fig:shadowgraph1} and \ref{fig:shadowgraph2}, and show that there is correlation between the decay of the trailing denser fluid and the entrainment timescale. 

Figures \ref{fig:shadowgraph1} and \ref{fig:shadowgraph2} showed a sequence of shadowgraph images for (A) a larger and (B) a smaller droplet, both composed of the densest oil mixture ($\rho_d$ = 0.9927 g/cm$^3$) and rising through a 4.6 cm transition region. The most obvious difference between these two cases is the asymmetry of the droplet's wake. The larger droplet shown in case A, with Re$_l = 370$, appears to be shedding vortices in a zigzag pattern, while the smaller droplet with Re$_l = 196$ has a highly symmetric wake structure. This aligns with the delineation of the effect of Reynolds number on the wakes of rising and falling spheres in a homogeneous fluid presented by Horowitz and Williamson \cite{horowitz_effect_2010}, who found that wake structures transition from vertical to oblique at Re $= 210$, and from oblique to zigzag at Re $ = 260$. The results are also qualitatively very similar to those for spheres in a linear stratification \cite{hanazaki_jets_2009}; in our experiments, the finite transition between two homogeneous-density fluids appears to constrain the wake structure to the transition region. While the vortex structures shed by the drop as it leaves the homogeneous lower layer vary significantly between these two experimental cases, the two cases have very similar Froude numbers (0.55 and 0.53), and indeed very similar entrainment and retention times (for example, $t_e = 4.95$ and 4.18 sec; $\tau_e = 29.5$ and 24.9 respectively). This variation in wake structure when entering and initially exiting the transition region thus appears to have little effect on fluid entrainment and drop retention.

\begin{figure}
  \centerline{\includegraphics[trim={2cm 5.5cm 1.7cm 6cm}, clip, width=.7\textwidth]{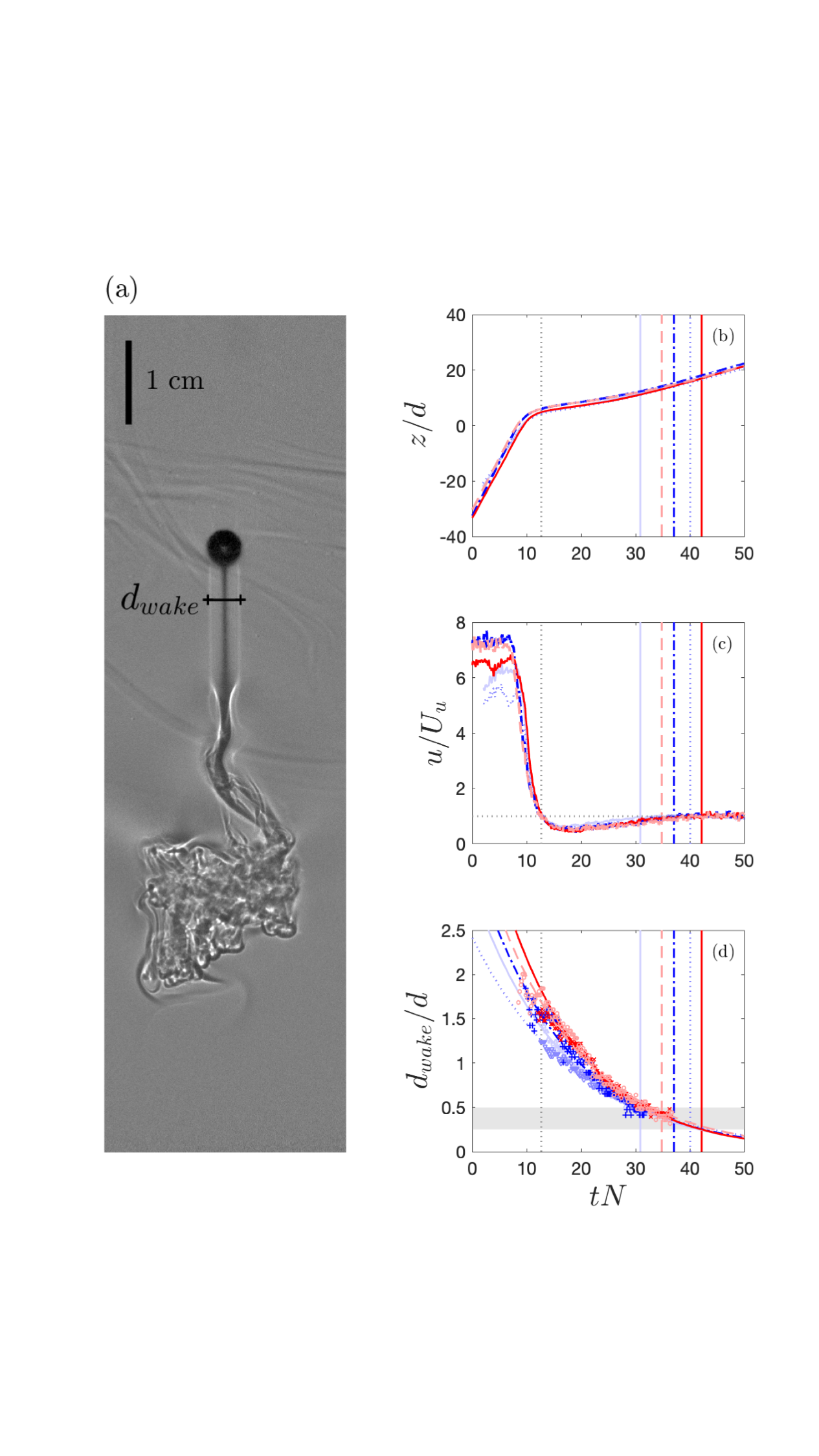}}
  \caption{Nondimensionalized tracking and wake diameter data for the shadowgraph experiments shown in figures \ref{fig:shadowgraph1} and \ref{fig:shadowgraph2} (red lines), as well as three other experiments (blue lines). Solid red lines correspond to Shadowgraph A, and dashed light red lines correspond to Shadowgraph B. (a) An example of wake diameter measurement. (b,c) Drop position and velocity versus time. The vertical dotted grey lines on (b-d) show when the drops first reach $U_u$; the length of time between this and the vertical lines of different shading indicate $\tau_e$ for each case. (d) The ratio of drop wake diameter to drop diameter, versus time. For all five cases, values of $d_{\text{wake}}/d$ were fit to an exponential curve (lines of corresponding color), yielding a decay coefficient of 0.60$\pm0.06$, with an average $R^2$ of 0.97.} 
\label{fig:d_wake}
\end{figure}

Although the far-field wake of the drop seems to have no impact, the gradual bleeding away of the local tail of fluid carried by the drop may instead play a role in drop retention. In figures \ref{fig:shadowgraph1} and \ref{fig:shadowgraph2}, the width of the tail of fluid dragged by the drop (denoted by changes in ambient illumination in the shadowgraph) slowly decreases over time. Although shadowgraphs are generally a qualitative tool for understanding density variations, we were able to estimate the approximate diameter of the trailing fluid carried by the drop as it rises through and past the transition region. These values were measured manually for the two cases shown here---as well as three others for which tracking and shadowgraph data were available---approximately one diameter below the bottom of the drop as it rises, as shown in figure \ref{fig:d_wake}(a). Because bright and dark regions of the shadowgraph indicate regions of strong concavity in the density field (i.e., large values of $\nabla^2 \rho$), actual perturbations in the density field persist slightly farther than can be observed in the shadowgraph. The estimated diameter is thus some fixed fraction of the actual wake diameter; however, we believe this is an adequate analog to examine trends in the wake over time. The ratio of wake diameter to drop diameter is plotted in figure \ref{fig:d_wake}(d) versus time nondimensionalized by the buoyancy frequency $N$. Nondimensionalized position and velocity for each case are also included in (b) and (c) for easy comparison. Time series of wake diameter are shorter than those of tracked position, as the shadowgraphs images were zoomed in closer to see fine details and the drop thus remained in the field of view of the camera for a shorter period of time.

An exponential decay can be fit to all five sets of data. Each time series was fit to a function of the form

\begin{equation}
    \frac{d_{\text{wake}}}{d} \propto \exp{\left(k N t\right)},
\end{equation}

\noindent yielding an average best-fit decay coefficient of $k = -0.60 \pm 0.06$ across all five cases, with an average $R^2$ of 0.97. The time at which the drops have asymptoted to their upper-layer terminal velocity, denoted by the vertical colored lines in figure \ref{fig:d_wake}(b-d), coincides with when the wake diameter is equal to between 1/4 and 1/2 of the drop diameter. Once the tail has become about two to four times smaller than the droplet, the drop reaches its homogeneous upper-layer behavior. The decay of this tail of denser fluid correlates well with the measured entrainment timescale for each case (i.e., the time between the vertical dotted grey lines in figure \ref{fig:d_wake}(b-d) and the vertical lines of different shading), and supports our interpretation of this timescale as an indicator of fluid entrainment. 


To our knowledge, this is the first experimental work to quantitatively measure the wake of a spheroid in a density transition, and relate it to the spheroid's slowdown and retention. While these visualization results are preliminary and limited to only a single drop density, we find that this is a promising path to explore further, and hope that a similar approach could be used to physically understand the force due to entrained fluid studied previously \cite{srdic-mitrovic_gravitational_1999,torres_flow_2000,yick_enhanced_2009,verso_transient_2019,zhang_core_2019}.


\subsection{Comparison to other work on drag enhancement in stratification}
\label{sec:dragenhancement}

While we have framed our observations above in terms of timescales, the experimental data from this study also allow for computation of enhanced drag forces in the presence of stratification. Here, we will present a theoretical formulation for this drag enhancement and compare it to the parameter space and force balances presented by Zhang et al. \cite{zhang_core_2019} for numerical studies of rigid spheres.

The forces acting on the particles or droplets passing through a stratified fluid are typically decomposed into several different terms \cite{clift_bubbles_2005,srdic-mitrovic_gravitational_1999,verso_transient_2019}:
\begin{align}
F_{tot} &= m \frac{\partial u}{\partial t} = F_b + F_d + F_a + F_h + F_s 
\end{align}
where $F_b$ is the buoyancy force, $F_d$ is the drag force, $F_a$ is the added mass force, $F_h$ is the Basset history force, and $F_s$ is any additional force attributable to interactions with the ambient stratification.
Here, we assume $F_b$ and $F_d$ correspond to the buoyancy and drag forces, respectively, which would be experienced by a droplet moving through the undisturbed ambient fluid. (Therefore, the density corresponds to the value for the undisturbed background fluid at the current location of the droplet.) Other work in a similar parameter regime has shown the added mass and history forces are typically much weaker than other forces in the system, and can be neglected \cite{verso_transient_2019}.

First, we will examine the functional form of the buoyancy and drag forces in a homogeneous fluid, $F_{bh}$ and $F_{dh}$, far away from the transition region. The buoyancy force can be computed as a function of the local fluid density $\rho_f$, which varies with depth $z$ \cite{srdic-mitrovic_gravitational_1999}:

\begin{equation}
    F_{bh}(\rho_f) = \frac{\pi}{6}\left(\rho_f - \rho_d\right) g d^3 = F_0 \text{Ar}_f,
    \label{eq:Fbh}
\end{equation}

\noindent where $F_0 = \frac{\pi \mu^2}{6 \rho_f}$. We can relate this buoyancy force to the local terminal velocity, $U_f(z)$, using the relationship measured in section \ref{sec:terminalvelocity},

\begin{equation}
    \text{Re}_f \approx \alpha \text{Ar}_f^{\beta},
\end{equation}

\noindent and the definition of Reynolds number, Re$_f = \rho_f U_f d/\mu$, and rearrange to obtain an expression for the Archimedes number in terms of the local droplet velocity, 


\begin{equation}
    \text{Ar}_f(U_f) = \left(\frac{U_f}{U_0}\right)^{1/\beta}
    \label{eq:Arh}
\end{equation}

\noindent where $U_0 = \frac{\alpha \mu}{\rho_f d}$, and the empirical measurements discussed in section \ref{sec:terminalvelocity} gave $\alpha = 0.60$ and $\beta = 0.57$.

In a homogeneous fluid, a droplet will reach its terminal velocity when $F_{dh} = -F_{bh}$. Using this relation and equations \ref{eq:Fbh} and \ref{eq:Arh} we can obtain a drag law for the drag force in a homogeneous fluid as

\begin{equation}
    F_{dh}\left(U_f\right) = -F_0 \left(\frac{U_f}{U_0}\right)^{1/\beta}.
    \label{eq:Fdh}
\end{equation}

We now use the relations given in equations \ref{eq:Fbh}, \ref{eq:Arh}, and \ref{eq:Fdh} in a simplified force balance in a stratified fluid. The force balance is considerably more tractable if we focus on the force enhancement at the point of minimal velocity ($t= t_{min}$, $u = U_{min}$), which is when the stratification force is at its maximum \cite{verso_transient_2019}. At this point, $\frac{\partial u}{\partial t}=0$, $F_a$ is a function of $\frac{\partial u}{\partial t}$ and so is also zero, we assume $F_h$ is negligible \cite{verso_transient_2019}, and the droplet is usually entirely in the upper fluid layer (see e.g. figures \ref{fig:shadowgraph1}, \ref{fig:shadowgraph2}, \ref{fig:z_vel_ex}). The force balance from equation \ref{eq:forcebalance} can thus be assumed to simplify to

\begin{equation}
    F_{s,max} = -F_b - F_d
    \label{eq:forcebalance}
\end{equation}

\noindent where $F_b$ and $F_d$ are the local buoyancy and drag forces when the droplet has just entered the upper layer, and is moving at its minimum velocity $U_{min}$.

This local buoyancy force is computed for a droplet that has fully entered the upper layer using equations \ref{eq:Fbh} and \ref{eq:Arh}:

\begin{equation}
    F_b = F_{bh}(\rho_u) = F_0 \text{Ar}_u = F_0 \left(\frac{U_u}{U_0}\right)^{1/\beta}.
    \label{eq:Fb}
\end{equation}

\noindent Using equation \ref{eq:Fdh}, the local drag force scales with the local velocity to the power of $(1/\beta)$,

\begin{equation}
    F_d = F_{dh}(U_{min}) = -F_0 \left( \frac{U_{min}}{U_0}\right)^{1/\beta}.
    \label{eq:Fd}
\end{equation}


Thus, we can write a drag enhancement ratio, $\Gamma$, which compares the maximum stratification force experienced by the droplet to the drag force:

\begin{equation}
    \Gamma = \frac{F_{s,max}}{F_d} = \frac{-F_b - F_d}{F_d} = \frac{-F_b}{F_d} - 1.
\end{equation}

\noindent Substituting the expressions from equations \ref{eq:Fb} and \ref{eq:Fd}, we find that

\begin{equation}
    \Gamma = \left(\frac{U_u}{U_{min}}\right)^{1/\beta} - 1.
\end{equation}

\begin{figure}
  \centerline{\includegraphics[width=.5\textwidth]{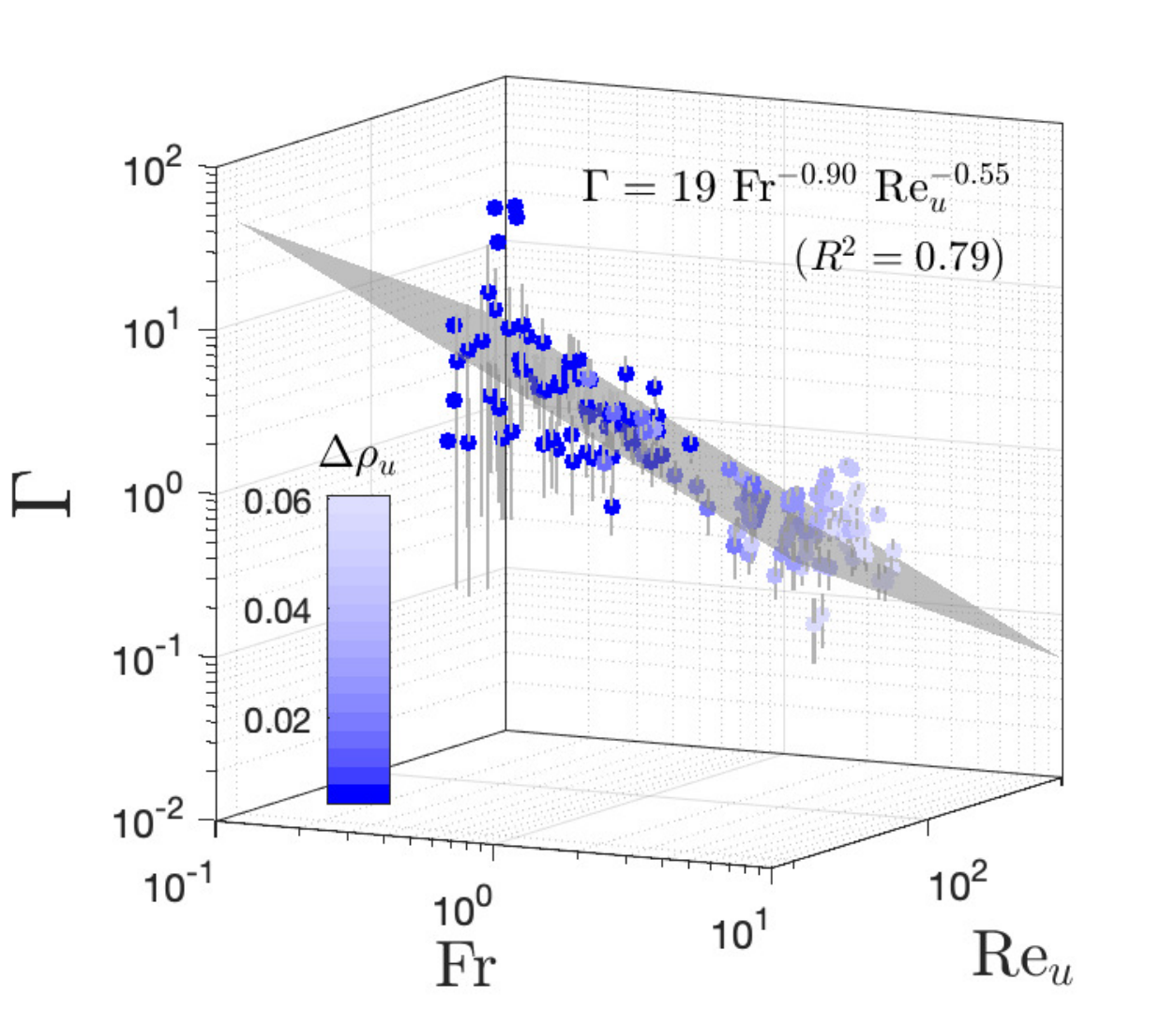}}
  \caption{The drag enhancement at the point of the velocity minimum, $\Gamma = \frac{F_{s,max}}{F_d}$, versus upper-layer Reynolds number and Froude number.}
\label{fig:gamma}
\end{figure}

This drag ratio $\Gamma$ is shown in figure \ref{fig:gamma} as a function of Re$_u$ and Fr. A power-law fit was found, with $\Gamma = 19 $ Fr$^{-0.90}$ Re$_u^{-0.55}$. These exponents agree remarkably well with the dependence of the drag force due to vorticity induced by baroclinic torque, $F_{\rho\omega}$, proposed by Zhang et al. \cite{zhang_core_2019}. In their Regimes 2 and 3 (Pr$^{-1/3} \ll$ Fr $\ll 1$ and Fr $\gg 1$, respectively, and in which all of our experimental measurements fall), they found that

\begin{equation}
    F_{\rho \omega} \sim \text{Fr}^{-1} \text{Re}^{-1/2}.
\end{equation}

\noindent In figure \ref{fig:gamma}, the largest errors in $\Gamma$ occur for the smallest Froude numbers, which fall towards the border of Regime 1 (for large Re and Pr, Regime 1 is delineated as Fr $\ll$ Pr$^{-1/6}$, or Fr $\ll 0.34$ in our case). Our smallest Froude number is 0.38. The scaling of drag enhancement from our experimental data therefore agrees best where we expect it to: farther from the edges of the regimes delineated in Zhang et al. \cite{zhang_core_2019}.


In summary, these experimental and analytical results regarding the drag enhancement due to stratification show that the baroclinic vorticity contribution is important in our experiments, and provide the first experimental data supporting the approach of Zhang et al \cite{zhang_core_2019}. Both of the relevant forces that contribute to ``stratification drag" discussed by Zhang et al. -- the force due to relative buoyancy of entrained fluid ($F_{\rho\rho}$), and the force due to modification of the local vorticity field due to baroclinic torque ($F_{\rho\omega}$) -- are predicted to scale with Fr$^{-1}$ in the parameter space covered by our experiments. Zhang et al. showed that $F_{\rho\omega}$ should dominate in the Reynolds number regime studied here, which results in our observed scaling of $\Gamma$ with both Fr and Re$_u$. We do expect $F_{\rho\rho}$ to play a role in this regime based on our timescale results and flow visualization of the droplet wake; however, its contribution is likely less than that of $F_{\rho\omega}$. It should be noted that the simulations of Zhang et al. were conducted in a linear stratification, so their definitions of Reynolds and Froude numbers vary from the definitions used in this study. Regardless, these experimental results demonstrate that the baroclinic vorticity force $F_{\rho \omega}$ is significant in the regime studied here.

\section{Discussion}
\label{sec:discussion}


We have studied the retention and entrainment dynamics of droplets in the regime $0.38 <$ Fr $< 4.2$, $59 <$ Re$_l < 1060$, $5.4 < $ Re$_u < 580$, with relative drop densities $\Delta \rho_u$ ranging from 0.0045 to 0.061, and $h/d$ ranging from 4.5 to 52. Counter to the results of Srdic-Mitrovic et al. \cite{srdic-mitrovic_gravitational_1999}, who found that experiments at higher Re$_u$ ($>15$) showed no significant change in drag as a sphere settled through a density gradient, we observed significant drop retention and slowdown for Re$_u$ up to 300, and for a wide range of Re$_l$. As noted in previous work \cite{abaid_internal_2004,verso_transient_2019}, the parameters of the layer that a spheroid is entering appear to be the most critical for observing ``levitation" or significant delay. This holds true in our study, in which Froude number and Reynolds numbers based on the upper layer terminal velocity are the governing parameters. 

Although our results regarding decay of a droplet's wake are promising, a more systematic study is required to draw stronger conclusions about the relationship between the entrainment timescale and the diameter, length, and overall shape of the trailing fluid column, as well as about the vortex shedding that happens before and during passage through the transition region. We also explored only a certain Froude and Reynolds number regime in this study; it remains to be seen whether this scaling applies to drops that are in the Stokes regime, or for extremely high Reynolds number spheroids, such as rising bubbles.

Finally, we have not systematically varied the Marangoni forces in this study. However, these forces are implicitly included in $\Gamma$, so any interfacial tension effects are captured in our measurement of the additional stratification-related forces. Our results regarding the scaling of drag enhancement with Reynolds and Froude number are comparable to the literature from solid particles, suggesting that this is a valid approach.

The dynamics explored in this study are applicable to a range of environmental scenarios, including oil spills and natural oil seeps. Although our results are for liquid droplets in an ambient stratification, we expect some of these findings to hold for solid particles as well, and may have applications in sediment suspension in benthic boundary layers \citep{adams_suspended-sediment_1981} and dispersal of pollutants in the atmosphere \citep{turco_climate_1990}.







\section{Conclusions}
\label{sec:conclusions}

In this study, we characterized the dynamics governing retention of a single droplet at a transition in density between two homogeneous fluids. We examined fluid flow and droplet retention for a range of drop sizes, drop densities, and ambient stratification profiles, allowing us to characterize drop behavior for a range of Reynolds and Froude numbers. We found that far from the density transition, within the homogeneous fluid layers, the droplets followed a balance between buoyancy, inertial, and viscous forces, and that Re$_f \sim$ Ar$_f^{\beta}$, where $\beta = 0.57$.

We explored two metrics measuring the timescale of drop delay at a density transition. The first metric, the entrainment time $t_e$, measures the amount of time that denser fluid is appreciably entrained, reducing the drop's speed. The second, the retention time $t_r$, measures the degree to which the droplet's rise is delayed. The retention time is related to the entrainment time by a simple linear relation involving the magnitude of the slowdown the droplet experiences, $t_r = 0.46 t_e (\Delta U/U_u)$, where $\Delta U = U_u - U_{min}$. The timescales and simple scaling arguments from our data lend themselves particularly well to experimental quantification of the trapping dynamics, as they are readily measurable from kinematic data and do not require taking derivatives or quantifying various forces at all points in time.


In the regime covered by our experiments ($0.38 <$ Fr $< 4.2$, $59 <$ Re$_l < 1060$, $5.4 < $ Re$_u < 580$, $0.0045 < \Delta \rho_u < 0.61$), nondimensional entrainment and retention times were found to depend on the Froude number and Reynolds number. Significant retention with either timescale only occurred for Fr $\lesssim 1$, suggesting that retention is primarily a function of the ratio of the buoyancy timescale ($1/N$) to the inertial timescale ($d/U_u$), and that trapping dynamics are dominated by the effects of stratification. We also found that the trailing column of dense fluid entrained by a droplet decays over time. The point at which the wake diameter becomes on the order of 1/4 to 1/2 of the drop diameter coincides with the drop reaching its upper-layer terminal velocity, supporting our interpretation of the entrainment timescale as the time required for denser trailing fluid to bleed away. This effect appears to be independent of the type of large-scale wake (zigzagging or vertical) the droplet has when first entering the transition region.

Finally, based on a force balance at the point of the droplet's minimum velocity, we derived a drag enhancement ratio $\Gamma = \frac{F_{s,max}}{F_d} = (U_u/U_{min})^{1/\beta} - 1$, and found a power-law fit to our experimental data of $\Gamma \sim$ Fr$^{-0.90}$Re$_u^{-0.55}$. This compares very favorably with the behavior of the baroclinic vorticity force proposed by Zhang et al., and suggests that drag forces generated by the baroclinic torque, likely in addition to the more traditionally-studied buoyancy-related entrainment forces, are important in this problem.

\begin{acknowledgments}
S.K. acknowledges support from the Hellman Faculty Fellows Fund. The authors thank Thomas Bellotti and Joshua Roe for contributing to numerous discussions on this project. The Photron cameras used in this study were generously lent by Jian-Qiao Sun. The authors also acknowledge the Joint Applied Math and Marine Sciences Fluids Lab at the University of North Carolina at Chapel Hill, where S.K. was a postdoctoral researcher, including Roberto Camassa, Richard McLaughlin, Holly Arrowood, Lauren Colberg, Elaine Monbureau, Sarah Spivey, and Arthur Wood for contributing to preliminary stages of this project. 
\end{acknowledgments}

\bibliography{oildroplets.bib}

\end{document}